\documentclass[prb,onecolumn,showpacs,preprintnumbers,amsmath,amssymb,floatfix]{revtex4}

\usepackage{enumerate}
\usepackage{graphicx}
\usepackage{subfigure}

\begin{document}

\title{Inhomogeneous backflow transformations in quantum Monte Carlo}

\author{P.\ L\'opez R\'{\i}os, A.\ Ma, N.~D.\ Drummond, M.~D.\ Towler,
and R.~J.\ Needs}

\affiliation{Theory of Condensed Matter Group, Cavendish Laboratory,
University of Cambridge, J.~J.~Thomson Avenue, Cambridge CB3 0HE,
United Kingdom}

\date{\today}

\begin{abstract}
An inhomogeneous backflow transformation for many-particle wave
functions is presented and applied to electrons in atoms, molecules,
and solids.  We report variational and diffusion quantum Monte Carlo
(VMC and DMC) energies for various systems and study the computational
cost of using backflow wave functions.  We find that inhomogeneous
backflow transformations can provide a substantial increase in the
amount of correlation energy retrieved within VMC and DMC
calculations.  The backflow transformations significantly improve the
wave functions and their nodal surfaces.
\end{abstract}

%

\maketitle


\section{Introduction}
\label{sec:introduction}

The \textit{fermion sign problem} continues to preclude the
application of in principle exact quantum Monte Carlo (QMC) methods
to large systems, and so approximate QMC methods must be used
instead.  Probably the most widely-used of these is the stable and efficient 
diffusion quantum Monte Carlo (DMC) algorithm,\cite{ceperley_1980,foulkes_2001}
in which the fermion sign problem is sidestepped through the introduction
of the fixed-node approximation.\cite{anderson_1975}
DMC can provide highly accurate energies for assemblies of quantum
particles, but the fixed-node approximation is uncontrolled and its
accuracy is often difficult to assess.

The fixed-node approximation\cite{anderson_1975} involves constraining the 
nodal surface of the wave function to equal that of an approximate ``trial''
or ``guiding'' wave function.  The fixed-node DMC energy is higher than the
ground-state energy, becoming equal in the limit that the fixed nodal surface
is exact.  The dependence of the DMC energy on the quality of
the trial wave function is often significant in practice.  It would therefore
be very useful to be able to construct trial wave functions with better nodal
surfaces to reduce the effect of the fixed-node approximation.

Efforts to construct wave functions with accurate nodal surfaces have
continued since the introduction of the fixed-node approximation.
Single-determinant wave functions often provide good nodal surfaces
for closed-shell systems, and multideterminant wave functions can do
so for small open-shell systems, although the required number of
determinants becomes excessive for large systems.  Compact pairing
wave functions consisting of an antisymmetrized product of
two-electron ``geminals''\cite{shull_1959} were introduced long
ago\cite{fock_1950,hurley_1953} and have recently been used in QMC
calculations for atoms and molecules.\cite{casula_2003,casula_2004}
Triplet-pairing Pfaffian wave functions were first used in QMC
calculations for liquid $^{3}$He by Bouchaud and
Lhuillier,\cite{bouchaud_1987} and recently this approach has been
extended by Bajdich \textit{et al.},\cite{bajdich_2006} who considered
atomic and molecular systems in which both parallel- and
antiparallel-spin electrons are paired.

Another approach for improving upon a single determinant of one-electron
orbitals is to introduce parameters which allow the orbitals to depend
on the positions of the other electrons.  Such a route was followed by
Wigner and Seitz,\cite{wigner_1934} who considered wave functions in
which the orbitals of the up-spin electrons depend on the positions
of the down-spin electrons, and \textit{vice versa}.  This idea surfaced again
much later in connection with the quantum-mechanical description of
``backflow.''  Classical backflow is related to the flow of a fluid
around a large impurity.  Its quantum analog was discussed by
Feynman\cite{feynman_1954} and Feynman and Cohen\cite{feynman_1956}
in the contexts of excitations in $^{4}$He and the effective mass of a
$^{3}$He impurity in liquid $^{4}$He.  They argued that the energy
would be lowered if the $^{4}$He atoms executed a flow pattern around
the moving $^{3}$He impurity which prevented the atoms overlapping
significantly.  This effect was shown to correspond to the requirement
that the local current of particles is conserved.  They recognized
that, without backflow, the effective mass of the $^{3}$He impurity
would equal the bare mass, and incorporating backflow led to a
substantial increase in the effective mass.  It turns out that the
mathematical form obtained by incorporating backflow into a
single-determinant wave function is related to the wave functions
considered by Wigner and Seitz.\cite{wigner_1934}

In later studies, wave functions including Jastrow factors and
backflow-like correlations were used to study a $^{3}$He impurity in
liquid $^{4}$He and liquid $^{3}$He within a Fermi-hypernetted chain
approximation.\cite{pandharipande_1973,schmidt_1979,manousakis_1983}
Backflow was first used in QMC calculations by Lee \textit{et
al.},\cite{lee_1981} who calculated the total energy of liquid $^{3}$He.
QMC calculations for electrons using Slater-Jastrow wave functions
with backflow correlations were first performed by Kwon \textit{et
al.}\cite{kwon_1993}\ for the two-dimensional homogeneous electron gas
(HEG), and later\cite{kwon_1998} for the three-dimensional HEG (see
also the paper by Zong \textit{et al.}\cite{zong_2002}).  QMC
calculations using Slater-Jastrow wave functions with backflow
correlations have also been reported for solid and liquid
hydrogen,\cite{holzmann_2003,pierleoni_2004} which were the first such
applications to inhomogeneous electron systems.

While Jastrow factors keep electrons away from one another and greatly
improve wave functions in general, they do not alter nodal surfaces.  Holzmann
\textit{et al.}\cite{holzmann_2003}\ have argued that backflow and
three-body Jastrow correlations arise as the next-order improvements
to the standard Slater-Jastrow wave function, which consists of a
Slater determinant multiplied by a two-body Jastrow factor.  The
importance of backflow correlations within DMC calculations is that
they alter the nodal surface and can therefore be used to
reduce the fixed-node error.

In this paper we introduce parameterized inhomogeneous backflow
transformations, and apply them to atoms, molecules, and extended
systems.  The rest of this paper is structured as follows: a general
description of the Slater-Jastrow and backflow wave functions is given
in Sec.~\ref{sec:sj}, an explicit form for the backflow displacement
field is developed in Sec.~\ref{sec:bf_transf}, an extensive set of
results is given in Sec.~\ref{sec:results} and discussed in
Sec.~\ref{sec:discuss}, and our conclusions are summarized in
Sec.~\ref{sec:conclude}.  Important technical information about the
calculations, including the constraints on the backflow parameters,
has been gathered in the appendices.  Hartree atomic units
($\hbar=|e|=m_e=4\pi \epsilon_0=1$) are used throughout.


\section{Slater-Jastrow and Slater-Jastrow-Backflow wave functions}
\label{sec:sj}

The Slater-Jastrow (SJ) wave function can be written as
\begin{equation}
\label{eq:psi_T_sj}
\Psi_{\rm T}^{\rm SJ}({\bf R})=e^{J({\bf R})} \Psi_{\rm S}({\bf R})
\;,
\end{equation}
where ${\bf R}$ denotes the set of electron coordinates $\{{\bf
r}_i\}$, $e^{J({\bf R})}$ is the Jastrow correlation factor, and the
Slater part $\Psi_{\rm S}({\bf R})$ consists of a determinant or sum
of determinants, defining the nodes of $\Psi_{\rm T}^{\rm
SJ}({\bf R})$.

Backflow (BF) correlations are introduced by substituting 
a set of \textit{collective coordinates} ${\bf X}$ for
the coordinates ${\bf R}$ in the Slater determinants, so that
\begin{equation}
\label{eq:psi_T_bf}
\Psi_{\rm T}^{\rm BF}({\bf R})=e^{J({\bf R})} \Psi_{\rm S}({\bf X})
\;,
\end{equation}
where each of the new coordinates is given
by\cite{lee_1981,schmidt_1981}
\begin{equation}
\label{eq:r2x}
{\bf x}_i={\bf r}_i+\mbox{\boldmath $\xi$}_i({\bf R})\;,
\end{equation}
where $\mbox{\boldmath $\xi$}_i$ is the backflow displacement of particle $i$, which
depends on the configuration of the whole system.


\section{Inhomogeneous backflow transformations}
\label{sec:bf_transf}

The form of the backflow displacement $\mbox{\boldmath $\xi$}_i$ in
\textit{homogeneous} systems has been taken
as\cite{lee_1981,schmidt_1981,kwon_1993}
\begin{equation}
\label{xi_ee}
\mbox{\boldmath $\xi$}_i^{\rm ee}=\sum_{j\neq i}^{N_e} \eta_{ij} {\bf r}_{ij} \;,
\end{equation}
where $N_e$ is the number of electrons and $\eta_{ij} = \eta(r_{ij})$ is
a function of the interparticle distance $r_{ij}$.
Eq.~(\ref{xi_ee}) is the most general isotropic two-electron
coordinate transformation for a homogeneous system.  A single electron
$i$ perceives space to be isotropic, but when another electron $j$ is
introduced, the electron-electron (e-e) vector ${\bf r}_{ij}$ becomes
an inequivalent direction.  The e-e backflow displacement is taken to
be along this direction, as there is no reason why a displacement in a
specific perpendicular direction should occur.

In a system with nuclei a new set of directions is introduced, the
electron-nucleus (e-n) vectors $\{{\bf r}_{iI}\}$, and one is led to
introduce an e-n contribution to $\mbox{\boldmath $\xi$}_i$, of the form
\begin{equation}
\mbox{\boldmath $\xi$}_i^{\rm en}=\sum_I^{N_n} \mu_{iI} {\bf r}_{iI} \;,
\end{equation}
where $\mu_{iI}=\mu(r_{iI})$ and $N_n$ is the number of nuclei.

We also introduce an electron-electron-nucleus (e-e-n) term to
describe two-electron backflow displacements in the presence of a
nearby nucleus,
\begin{equation}
\mbox{\boldmath $\xi$}_i^{\rm een}=\sum_{j\neq i}^{N_e} \sum_I^{N_n} {\big (}
\Phi_i^{jI} {\bf r}_{ij} + \Theta_i^{jI} {\bf r}_{iI} {\big )} \;,
\end{equation}
where $\Phi_i^{jI}=\Phi^I(r_{iI},r_{jI},r_{ij})$ and
$\Theta_i^{jI}=\Theta^I(r_{iI},r_{jI},r_{ij})$.  Note that the vector
$\Phi_i^{jI} {\bf r}_{ij} + \Theta_i^{jI} {\bf r}_{iI}$ is capable of
spanning the plane defined by ${\bf r}_i$, ${\bf r}_j$, and ${\bf
r}_I$, without the need to introduce a component along the direction
of ${\bf r}_{jI}$.  The total backflow displacement is the sum of
these three components, $\mbox{\boldmath $\xi$}_i=\mbox{\boldmath $\xi$}_i^{\rm ee}+
\mbox{\boldmath $\xi$}_i^{\rm en}+\mbox{\boldmath $\xi$}_i^{\rm een}$.

At large distances $\eta(r_{ij})$ is expected to decay as $r_{ij}^{-3}$
in three dimensions\cite{kwon_1998} and $r_{ij}^{-5/2}$ in two
dimensions.\cite{kwon_1993} However, for computational efficiency and
for compatibility with periodic boundary conditions it is better to
cut off the $\eta$ function and the other backflow functions smoothly
at some radius.  We use a simple cutoff function,
\begin{equation}
\label{eq:truncf}
f(r;L)={\bigg (}\frac{L-r}{L}{\bigg )}^C H(L-r) \;,
\end{equation}
where $r$ is to be substituted by an e-e or e-n distance as
appropriate, $L$ is the cutoff length, $C$ is the \textit{truncation
order},\footnote{The $C$-th derivative of the wave function will be
discontinuous at $r=L$.  In particular, its Laplacian, used in the
computation of the kinetic energy, is discontinuous at $r=L$ if $C<3$.
In this work we have only considered $C=2$ and $C=3$.} and $H$ denotes
the Heaviside function.  The advantages of this cutoff function are, firstly,
that its value can be computed rapidly and, secondly, that it has considerable
flexibility because one can choose the value of $C$ and use $L$ as an
optimizable parameter.

Rational\cite{kwon_1993} and Gaussian\cite{holzmann_2003} forms for
homogeneous backflow functions have been used in previous work.
However, we have chosen to use natural power expansions because of the
excellent results we have obtained with such expansions for our
Jastrow factor,\cite{drummond_2004} and the lack of \textit{a priori}
knowledge of more specific parameterizations for the inhomogeneous
functions.  It is estimated that numerical errors in the evaluation of
natural polynomials become significant beyond order about $20$ when
using double-precision arithmetic and, although one can go to
substantially larger orders using Chebyshev polynomials, we have not
found this to be an issue in our work.

We have used the following polynomial expansions for $\eta_{ij}$,
$\mu_{iI}$, $\Phi_i^{jI}$, and $\Theta_i^{jI}$,
\begin{eqnarray}
\label{eq:eta_poly}
\eta_{ij} & = & f(r_{ij};L_\eta) \sum_{k=0}^{N_\eta}c_k r_{ij}^k, \\
\label{eq:mu_poly}
\mu_{iI} & = & f(r_{iI};L_{\mu,I}) \sum_{k=0}^{N_{\mu,I}}d_{k,I}
r_{iI}^k, \\
\label{eq:Phi}
\Phi_i^{jI} & = & f(r_{iI};L_{\Phi,I}) f(r_{jI};L_{\Phi,I})
\sum_{k=0}^{N_{{\rm en},I}} \sum_{l=0}^{N_{{\rm en},I}}
\sum_{m=0}^{N_{{\rm ee},I}} \varphi_{klm,I} r_{iI}^k r_{jI}^l
r_{ij}^m, \\
\label{eq:Theta}
\Theta_i^{jI} & = & f(r_{iI};L_{\Phi,I}) f(r_{jI};L_{\Phi,I})
\sum_{k=0}^{N_{{\rm en},I}} \sum_{l=0}^{N_{{\rm en},I}}
\sum_{m=0}^{N_{{\rm ee},I}} \theta_{klm,I} r_{iI}^k r_{jI}^l r_{ij}^m,
\end{eqnarray}
where $N_\eta$, $N_{\mu,I}$, $N_{{\rm en},I}$, and $N_{{\rm ee},I}$
are the expansion orders, $L_\eta$, $L_{\mu,I}$, and $L_{\Phi,I}$, are
cutoff lengths, and $\{c_{k}\}$, $\{d_{k,I}\}$,
$\{\varphi_{klm,I}\}$, and $\{\theta_{klm,I}\}$ are the optimizable
parameters.  We allow the parameters in $\eta$, $\Phi^I$, and
$\Theta^I$ to depend on the spins of the electron pairs, and those in
$\mu$ to be spin dependent; for simplicity we have omitted such
dependencies in the description of the functional forms above.  In periodic
systems, we constrain $L_\eta$ and $L_\mu$ to be smaller than the
Wigner-Seitz radius $L_{\rm WS}$ of the simulation cell and $L_\Phi$
to be smaller than $L_{\rm WS}/2$, for computational efficiency.

\section{Results}
\label{sec:results}

In this section we present variational quantum Monte Carlo (VMC)
and DMC results obtained with our
implementation of backflow transformations.  The \textsc{casino}
code\cite{casino} has been used for all of our QMC calculations.  
Our DMC algorithm is essentially as described in
Ref.\ \onlinecite{umrigar_1993}.  All DMC energies reported here have
been extrapolated to zero time step.  We have optimized
the parameters in our wave functions by minimizing the unreweighted
variance of the energy,\cite{umrigar_1988} using a scheme which
facilitates the optimization of parameters that modify the nodal
surface.\cite{kent_1999,drummond_2005}

We have used the Jastrow correlation factor of Drummond \textit{et
al}.\cite{drummond_2004} In our all-electron (AE) calculations, with
the exception of those for the HEG, the orbitals were obtained from Hartree-Fock
(HF) calculations using large Gaussian basis sets and the
\textsc{crystal98} code,\cite{crystal98} and the cusp-correction
algorithm of Ref.\ \onlinecite{ma_2005} was applied to each orbital at
each nucleus.  In our pseudopotential (PP) calculations we used the Dirac-Fock Average Relativistic Effective
pseudopotentials of
Refs.\ \onlinecite{trail_ppots1} and \onlinecite{trail_ppots2}, the nonlocal energies
being calculated within the locality
approximation.\cite{mitas_1991}  The
one-electron orbitals were obtained from the plane-wave PP
\textsc{castep} code\cite{segall_2002} using the
Perdew-Burke-Ernzerhof (PBE)
generalized-gradient-approximation\cite{perdew_1996}
exchange-correlation functional.  The orbitals were re-expanded
in terms of ``blip'' functions,\cite{alfe_2004} making the QMC
calculations much more efficient.

We have reported the variance of the \textit{total} local energy for
our VMC calculations, $\sigma^2=\langle\hat H^2\rangle-
\langle\hat H\rangle^2$,\footnote{$\hat H$ is the Hamiltonian operator
of the system, and the total local energy is defined as $E_L=\Psi_T^{-1}
\hat H\Psi_T$.} while the reported mean energies are either total, per
electron, or per primitive cell, as we have found appropriate in each
case.  We have estimated the amount of correlation energy retrieved
in our calculations by comparing our energies with ``exact'' reference data,
where available.  In the case of the PP carbon atom and PP carbon dimer
we have used the estimates of the PP valence correlation energy of
Ref.~\onlinecite{dolg_1996} assuming an error bar of $0.004$~a.u. as
suggested by the author.  In the HEG we have used our
BF-DMC energies as if they were ``exact'', and in PP carbon diamond
we have not estimated the amount of correlation energy retrieved.


\subsection{Homogeneous electron gas}

We studied three-dimensional, unpolarized HEGs consisting of 54
electrons in a simple cubic simulation cell subject to periodic
boundary conditions.  As well as the densities of $r_s=1$, 5, 10, and
20 studied by Kwon \textit{et al.}\cite{kwon_1998}\ and Holzmann
\textit{et al.}\cite{holzmann_2003}\ using backflow wave functions, for
completeness we studied two additional densities, $r_s=0.5$ and
$2$.  Holzmann \textit{et al.}\ used an analytical backflow form containing
no variable parameters in addition to a Gaussian form with variable
parameters.  In each case we compare our result with the corresponding
lowest-energy backflow result from table II of
Ref.~\onlinecite{holzmann_2003}.

\begin{figure}[!ht]
\begin{center}
\includegraphics[scale=0.5]{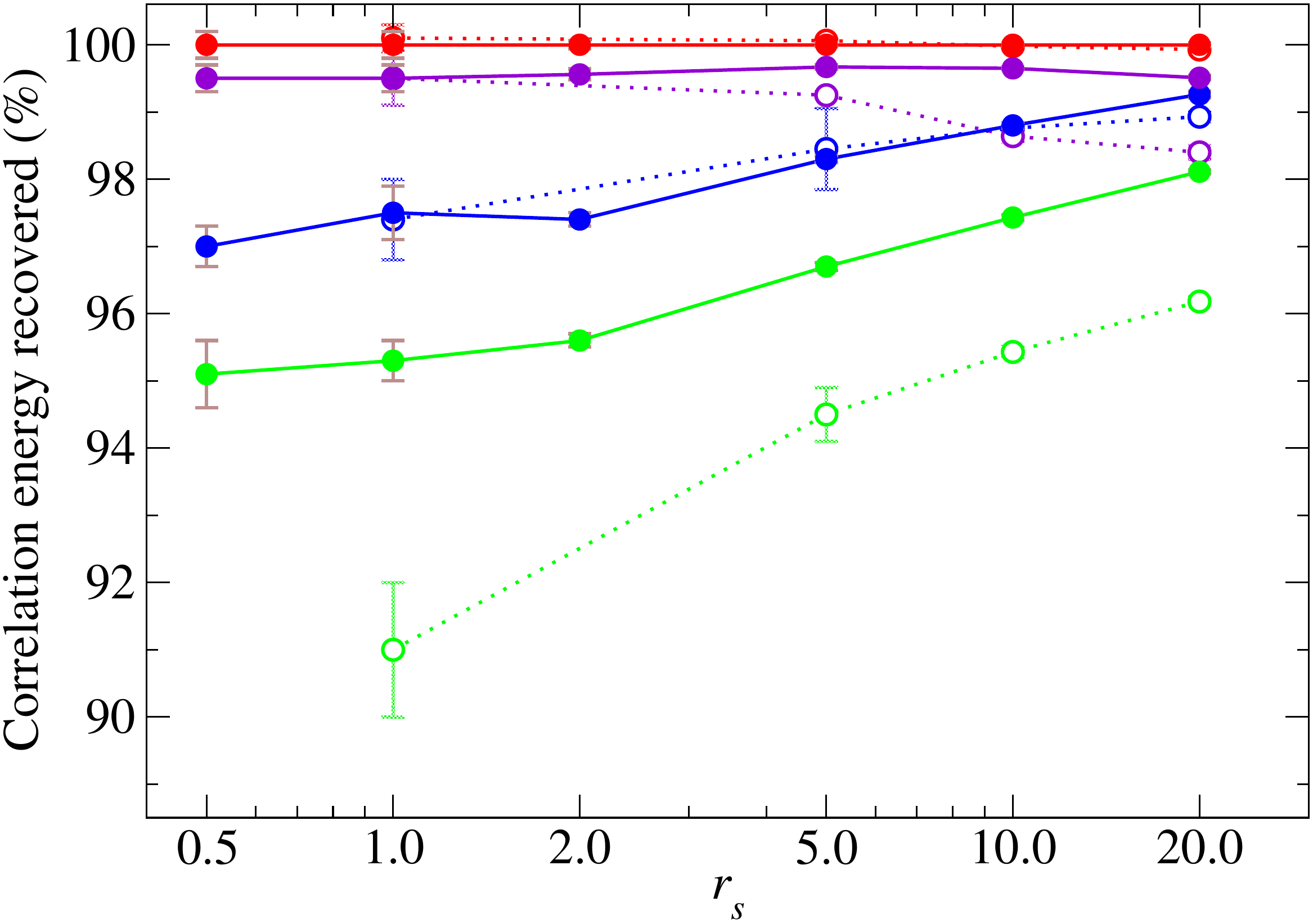}
\end{center}
\caption{(Color online) Percentages of the correlation energy
recovered at the (solid circles from top to bottom) BF-DMC, BF-VMC,
SJ-DMC, and SJ-VMC levels as a function of the density parameter $r_s$
(see Table~\ref{table:heg}).  Zero correlation energy corresponds to
HF-VMC and 100\% to BF-DMC\@.  The hollow circles are the best
BF-DMC, BF-VMC, SJ-DMC, and SJ-VMC energies of Holzmann \textit{et
al.},\cite{holzmann_2003} in the same order.  The statistical error
bars on the QMC data are smaller than the symbols except where
error bars are shown.
\label{fig:heg_ce}}
\end{figure}

We included a plane-wave term in our Jastrow factor [Eq.~(28) of
Ref.\ \onlinecite{drummond_2004}], which we found to improve the variational
energies at all densities.  We also studied the
effect of including a symmetric three-electron Jastrow term, $W$, of
the type used in Ref.\ \onlinecite{kwon_1993}, with
\begin{equation}
\label{eq:full_W_3}
W = \sum_i^{N_e} \sum_{j(\neq i)}^{N_e} \sum_{k(\neq i,j)}^{N_e}
    \left( w_{ij} {\bf r}_{ij} \right) \cdot
    \left( w_{ik} {\bf r}_{ik} \right) \;,
\end{equation}
where $w_{ij}$ is a function of the distance between electrons $i$ and
$j$, which we parameterized as
\begin{equation}
\label{eq:wij}
w_{ij} = f(r_{ij};L_w) \sum_{l=0}^{N_w} e_l r_{ij}^l \;,
\end{equation}
where $N_w$ is the order of the expansion, the $\{e_l\}$ are expansion
parameters, and $f$ is the cutoff function of Eq.~(\ref{eq:truncf}).
We decided to include a $W$ term for all densities at the
Slater-Jastrow level, while we used it in conjuction with backflow only
for the three lowest densities, where its effect on the SJ energy
was found to be statistically significant.  We refer to the SJ and BF wave
functions with a three-electron Jastrow term as SJ3 and BF3,
respectively.  The backflow parameters were allowed to depend on the
spins of the electron pairs, while the parameters in the
three-electron Jastrow factors were constrained to be independent of
spin, as this gave slightly better results.  The expansion orders
$N_\eta$ and $N_w$ were set to $8$ for all densities.  The cutoff
lengths $L_\eta$ and $L_w$ were optimized, but at all densities they
adjusted themselves to the maximum allowed value (the Wigner-Seitz radius).
The energies and variances obtained are given in Table~\ref{table:heg},
and the energies are illustrated in
Fig.~\ref{fig:heg_ce}, which gives the percentage of the correlation
energy retrieved at different levels as well as the SJ and BF energies
of Ref.\ \onlinecite{holzmann_2003}.  The introduction of backflow
increases the kinetic energy, but decreases the potential energy by a
larger amount.  Our SJ-DMC energies are in good agreement with those
of Holzmann \textit{et al.}, which of course they should be, because
the SJ trial wave functions have identical nodal surfaces.  Our SJ-VMC
calculations retrieve a higher percentage of the correlation energy
than those of Holzmann \textit{et al.}, and we believe this is mainly
due to the plane-wave term in our Jastrow factor.  Our BF-VMC calculations
consistently retrieve $99.5\%$ of the correlation energy throughout
the density range considered, while those of Holzmann \textit{et al.}\
drop below $99\%$ for $r_s>5$.  Our BF-DMC energies are within error
bars of those of Holzmann \textit{et al.}  In agreement with the work
of Refs.\ \onlinecite{holzmann_2003} and \onlinecite{kwon_1998},
we found that backflow
gives a larger energy reduction at the VMC level than the three-body
Jastrow term $W$ at all densities, although $W$ becomes more
important at large $r_s$.

\begin{figure}[!ht]
\begin{center}
\includegraphics[scale=0.5]{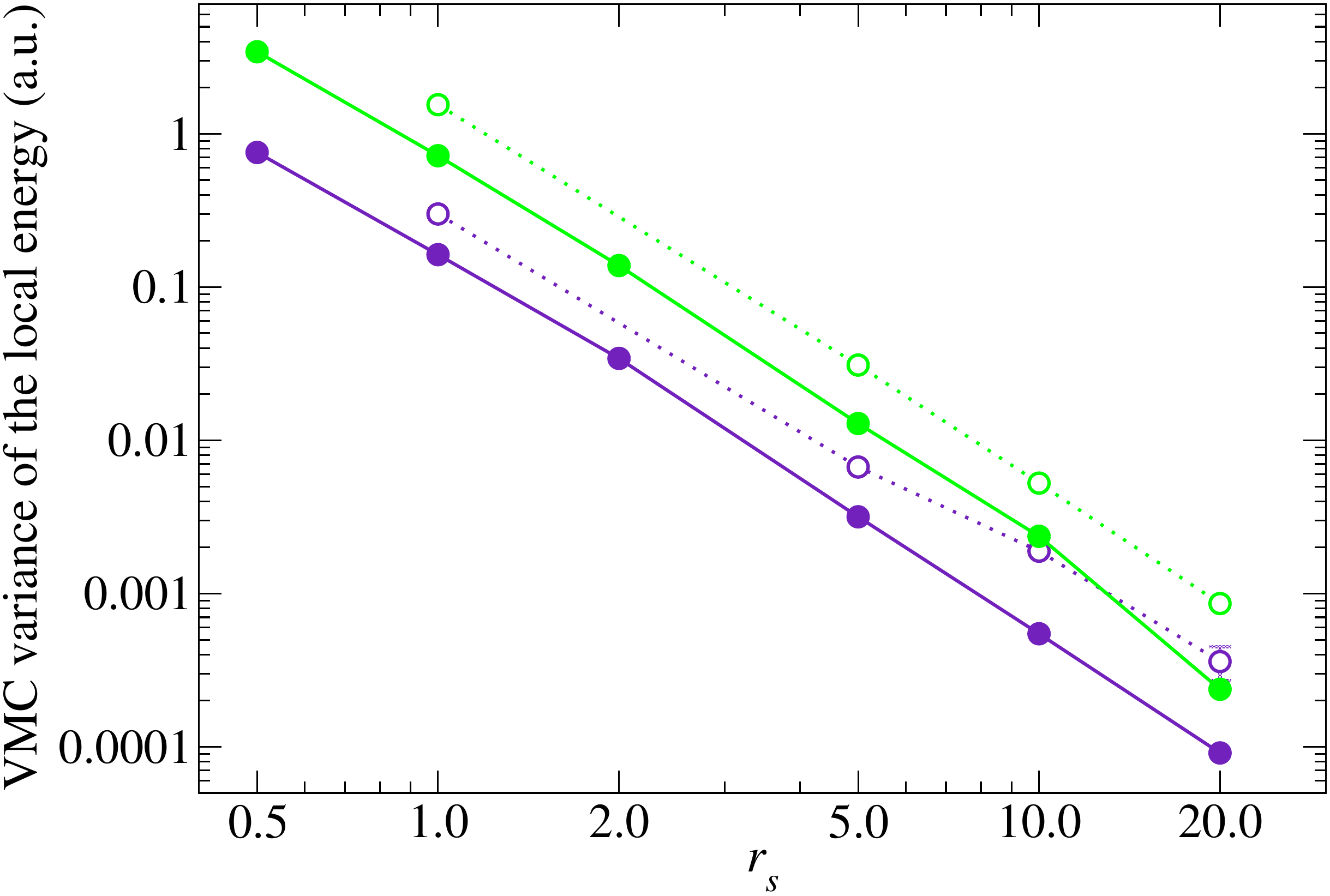}
\end{center}
\caption{(Color online) VMC variances achieved at the SJ level (top
solid line) and BF level (bottom solid line) as a function of the
density parameter $r_s$ (see Table~\ref{table:heg}).  The hollow
circles are the best SJ-VMC and BF-VMC variances of Holzmann
\textit{et al.},\cite{holzmann_2003} in the same order.  The statistical
error bars on the QMC data are smaller than the symbols except where
error bars are shown.
\label{fig:heg_var}}
\end{figure}

The variances of the VMC energies reported in Table~\ref{table:heg}
are illustrated in Fig.~\ref{fig:heg_var}.  The lines on the log-log
plot corresponding to our SJ and BF variances are almost parallel,
indicating an almost constant ratio of the SJ to BF variances of about
$4$.  The variances of Holzmann \textit{et al.}\  are systematically
higher than ours for comparable calculations, and at $r_s=20$
our SJ variance is lower than their BF variance.\footnote{The VMC
variances for the HEG at $r_s=1$ reported in Table II of Kwon
\textit{et al.}\cite{kwon_1998}\ have been confirmed by the authors to
be in error; the true values are a factor of $10$ smaller.  These data
were later used in Table II of Holzmann \textit{et
al.},\cite{holzmann_2003} who corrected the mistakes, except for the
variance of the SJ calculation.  We have compared our data with the
corrected values.}

\begin{table}[!ht]
\begin{center}
\begin{tabular}{r@{.}llr@{.}lr@{}lr@{}lr@{.}lr@{.}l}
\hline \hline
\multicolumn{2}{c}{$r_s$} & Wfn. &
\multicolumn{2}{c}{$E_{\rm V}$ (a.u./elec)} &
\multicolumn{2}{c}{$\sigma_{\rm V}^2$ (a.u.)} &
\multicolumn{2}{c}{${\rm CE}_{\rm V}$ (\%)} &
\multicolumn{2}{c}{$E_{\rm D}$ (a.u./elec)} &
\multicolumn{2}{c}{${\rm CE}_{\rm D}$ (\%)} \\
\hline
$0$&$5$  & HF  & $3$&$2659(7)$     & $76$&$(1)$       & $0$&$(3)$  &
             \multicolumn{2}{c}{-}                    & \multicolumn{2}{c}{-} \\
\multicolumn{2}{c}{}
         & SJ  & $3$&$2236(2)$     & $3$&$.34(3)$     & $94$&$.5(5)$  &
                 $3$&$22245(9)$                       & $97$&$0(3)$ \\
\multicolumn{2}{c}{}
         & SJ3 & $3$&$2233(2)$     & $3$&$.4(2)$      & $95$&$.1(5)$  &
             \multicolumn{2}{c}{-}                    &\multicolumn{2}{c}{-} \\
\multicolumn{2}{c}{}
         & BF  & $3$&$22132(7)$    & $0$&$.76(1)$     & $99$&$.5(2)$  &
                 $3$&$22112(4)$                       &$100$&$0(2)$
\vspace{0.2cm}\\
$1$&$0$  & HF  & $0$&$5689(4)$     & $19$&$.1(4)$     & $0$&$(2)$  &
             \multicolumn{2}{c}{-}                    &\multicolumn{2}{c}{-} \\
\multicolumn{2}{c}{}
         & SJ  & $0$&$53211(7)$    & $0$&$.719(7)$    & $94$&$.3(3)$  &
                 $0$&$53089(9)$                       & $97$&$5(4)$ \\
\multicolumn{2}{c}{}
         & SJ3 & $0$&$53175(7)$    & $0$&$.80(6)$     & $95$&$.3(3)$  &
             \multicolumn{2}{c}{-}                    &\multicolumn{2}{c}{-} \\
\multicolumn{2}{c}{}
         & BF  & $0$&$53009(3)$    & $0$&$.163(2)$    & $99$&$.5(2)$  &
                 $0$&$52989(4)$                       &$100$&$0(2)$
\vspace{0.2cm}\\
$2$&$0$  & HF  & $0$&$0186(2)$     & $4$&$.9(1)$      & $0$&$(1)$  &
             \multicolumn{2}{c}{-}                    &\multicolumn{2}{c}{-} \\
\multicolumn{2}{c}{}
         & SJ  & $-0$&$01246(3)$   & $0$&$.147(2)$    & $95$&$.4(1)$  &
                 $-0$&$01311(2)$                      & $97$&$4(1)$ \\
\multicolumn{2}{c}{}
         & SJ3 & $-0$&$01252(3)$   & $0$&$.138(2)$    & $95$&$.6(1)$  &
             \multicolumn{2}{c}{-}                    &\multicolumn{2}{c}{-} \\
\multicolumn{2}{c}{}
         & BF  & $-0$&$01382(2)$   & $0$&$.0342(6)$   & $99$&$.56(7)$ &
                 $-0$&$013966(9)$                     &$100$&$00(6)$
\vspace{0.2cm}\\
$5$&$0$  & HF  & $-0$&$05625(7)$   & $0$&$.76(1)$     & $0$&$.0(6)$   &
             \multicolumn{2}{c}{-}                    &\multicolumn{2}{c}{-} \\
\multicolumn{2}{c}{}
         & SJ  & $-0$&$07815(1)$   & $0$&$.0149(2)$   & $96$&$.09(7)$ &
             \multicolumn{2}{c}{-}                    &\multicolumn{2}{c}{-} \\
\multicolumn{2}{c}{}
         & SJ3 & $-0$&$078284(9)$  & $0$&$.0129(3)$   & $96$&$.70(6)$ &
                 $-0$&$078649(7)$                     & $98$&$30(5)$ \\
\multicolumn{2}{c}{}
         & BF3 & $-0$&$078961(5)$  & $0$&$.00317(6)$  & $99$&$.67(3)$ &
                 $-0$&$079036(3)$                     &$100$&$00(3)$
\vspace{0.2cm}\\
$10$&$0$ & HF  & $-0$&$03884(4)$   & $0$&$.194(4)$    & $0$&$.0(5)$   &
             \multicolumn{2}{c}{-}                    &\multicolumn{2}{c}{-} \\
\multicolumn{2}{c}{}
         & SJ  & $-0$&$053927(4)$  & $0$&$.00236(2)$  & $96$&$.69(4)$ &
             \multicolumn{2}{c}{-}                    &\multicolumn{2}{c}{-} \\
\multicolumn{2}{c}{}
         & SJ3 & $-0$&$054042(4)$  & $0$&$.00179(3)$  & $97$&$.43(4)$ &
                 $-0$&$054255(3)$                     & $98$&$80(4)$\\
\multicolumn{2}{c}{}
         & BF3 & $-0$&$054389(2)$  & $0$&$.00055(1)$  & $99$&$.65(2)$ &
                 $-0$&$054443(2)$                     & $100$&$00(2)$
\vspace{0.2cm}\\
$20$&$0$ & HF  & $-0$&$02205(2)$   & $0$&$.0477(9)$   & $0$&$.0(4)$   &
             \multicolumn{2}{c}{-}                    &\multicolumn{2}{c}{-} \\
\multicolumn{2}{c}{}
         & SJ  & $-0$&$031767(2)$  & $0$&$.000377(4)$ & $97$&$.20(4)$ &
             \multicolumn{2}{c}{-}                    &\multicolumn{2}{c}{-} \\
\multicolumn{2}{c}{}
         & SJ3 & $-0$&$031858(1)$  & $0$&$.000237(2)$ & $98$&$.11(3)$ &
                 $-0$&$031973(3)$                     & $99$&$26(5)$ \\
\multicolumn{2}{c}{}
         & BF3 & $-0$&$0319984(8)$ & $0$&$.000091(1)$ & $99$&$.51(3)$ &
                 $-0$&$032047(2)$                     & $100$&$00(3)$\\
\hline \hline
\end{tabular}
\caption{Energies and variances for three-dimensional, unpolarized HEGs
consisting of 54 electrons in a simple cubic simulation cell.  $E_{\rm
V}$ and $E_{\rm D}$ refer to VMC and DMC energies, respectively; ${\rm
CE}_{\rm V}$ and ${\rm CE}_{\rm D}$ are the percentages of the
correlation energies retrieved at the VMC and DMC levels,
respectively, and $\sigma_{\rm V}^2$ is the VMC variance.
\label{table:heg}}
\end{center}
\end{table}

\begin{figure}[!ht]
\begin{center}
\includegraphics[scale=0.5]{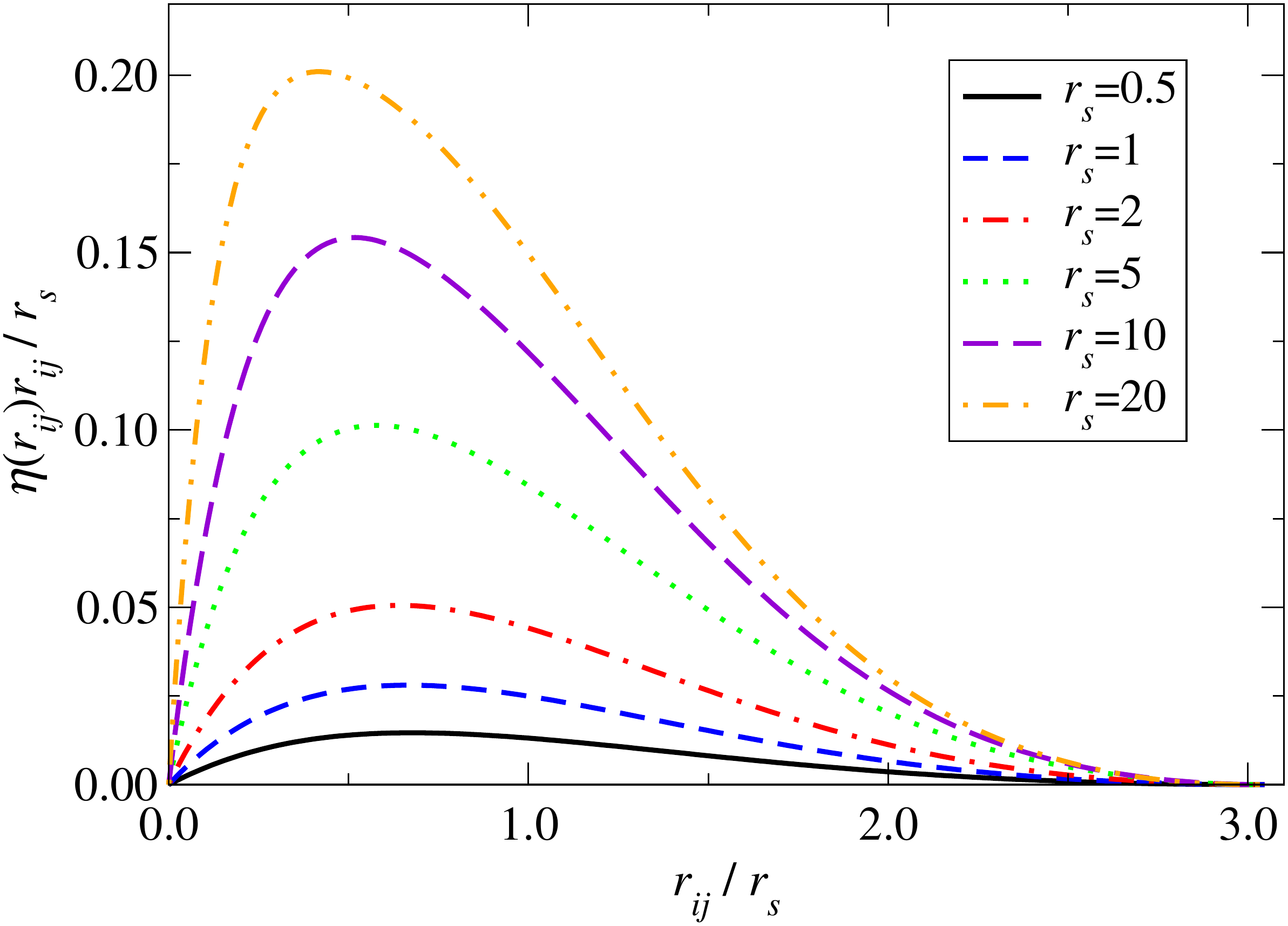}
\end{center}
\caption{(Color online) Antiparallel-spin homogeneous backflow
displacement $\eta(r_{ij})r_{ij}$ for the HEG at the
different densities studied.  For the three highest densities, the
curves correspond to BF wave functions, while the others are for
BF3 wave functions.
\label{fig:heg_eta_AP}}
\end{figure}

\begin{figure}[!ht]
\begin{center}
\includegraphics[scale=0.5]{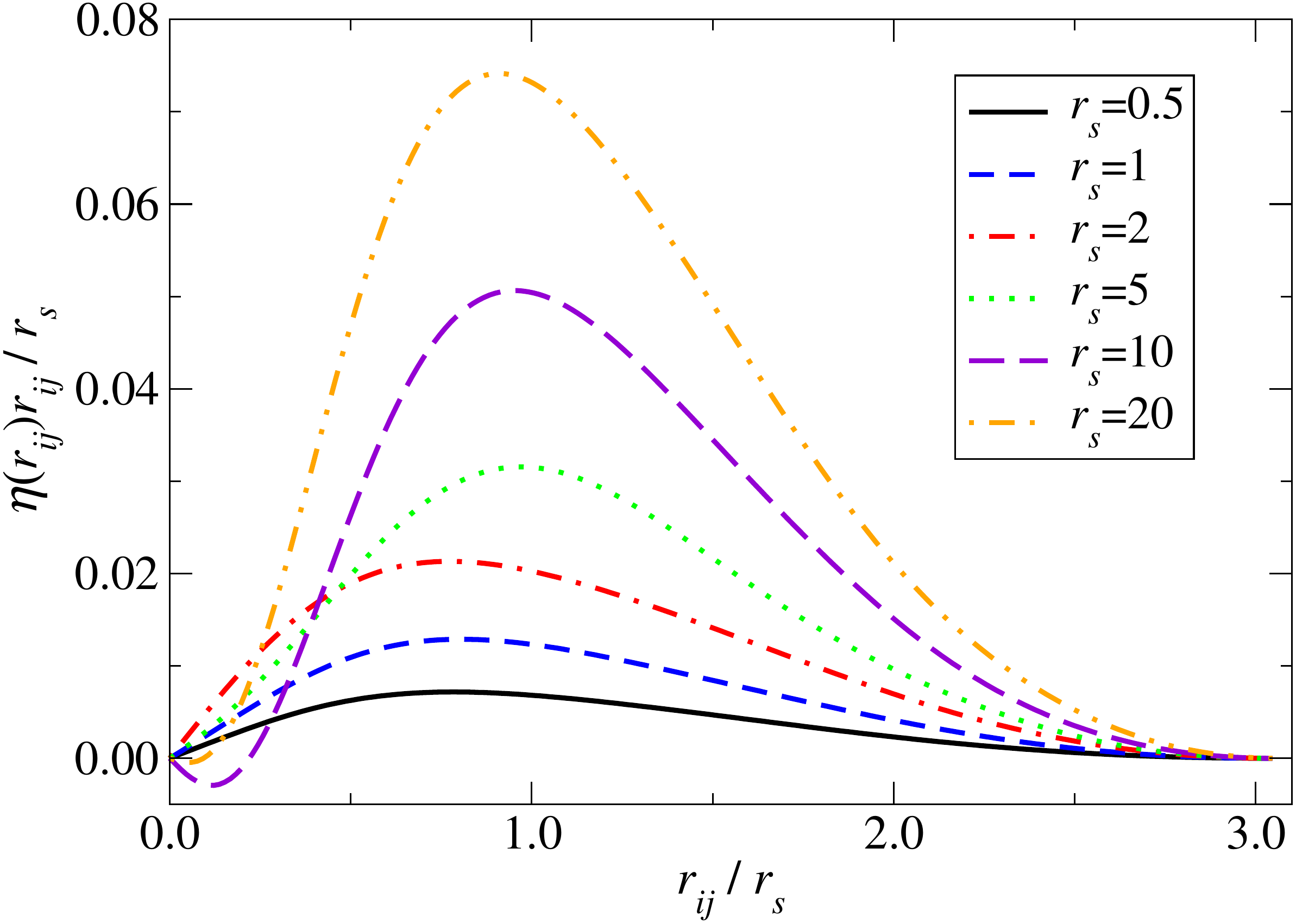}
\end{center}
\caption{(Color online) Parallel-spin homogeneous backflow
displacement $\eta(r_{ij})r_{ij}$ for the HEG at the different
densities studied.  For the three highest densities, the curves
correspond to BF wave functions, while the others are for the BF3
wave function.
\label{fig:heg_eta_P}}
\end{figure}

\begin{figure}[!ht]
\begin{center}
\includegraphics[scale=0.5]{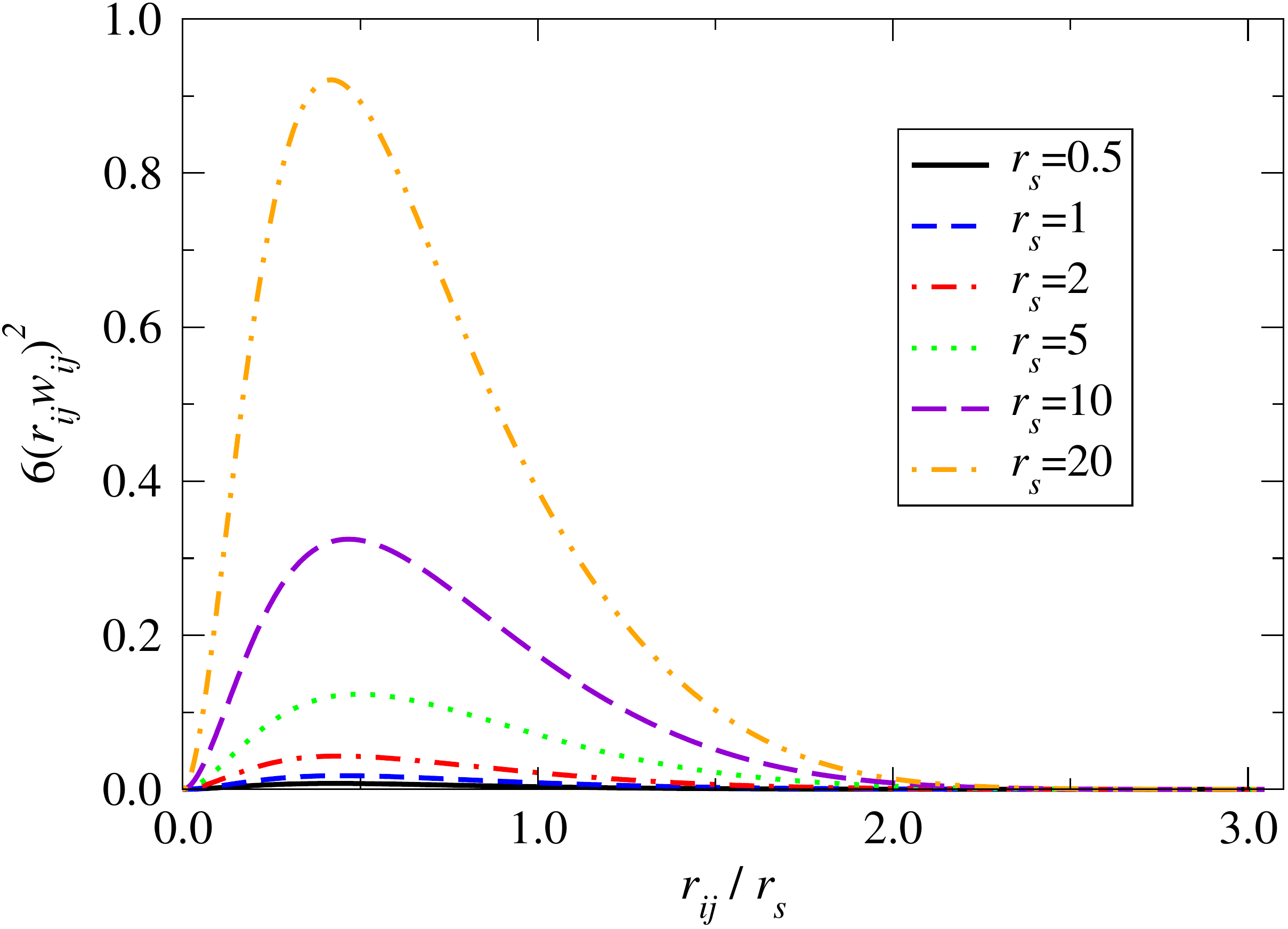}
\end{center}
\caption{(Color online) Three-body contribution to the Jastrow
function for the HEG due to three electrons at the vertices of an
equilateral triangle of side $r_{ij}$, at the different densities
studied.  For the three highest densities, the curves correspond
to SJ3 wave functions, while the other three are for BF3 wave
functions.
\label{fig:heg_w}}
\end{figure}

The optimized homogeneous backflow displacement $\eta(r_{ij})r_{ij}$
is plotted in Figs.~\ref{fig:heg_eta_AP} and \ref{fig:heg_eta_P}, and
the optimized three-body function $w_{ij}$ is shown in
Fig.~\ref{fig:heg_w}.  Holzmann \textit{et al.}\cite{holzmann_2003}\
and Kwon \textit{et al.}\cite{kwon_1998}\ used identical $\eta$ functions
for parallel and antiparallel spin pairs, whereas we have allowed them
to differ.  At each density, the maximum value of $\eta(r_{ij})r_{ij}$
for antiparallel spins is over twice as large as that for parallel
spins, and occurs at smaller electron separations.  The backflow
displacements for antiparallel spins are generally larger than for
parallel spins, and hence antiparallel-spin backflow is much more important
than parallel-spin backflow.  Our antiparallel-spin $\eta$ function is
similar to the spin-independent $\eta$ function of Kwon \textit{et
al.},\cite{kwon_1998} except that we do not find an attractive tail at
$r_s=20$. Note that, to obey the cusp conditions, we constrain the
parallel-spin $\eta(r_{ij})$ function to have zero derivative at
$r_{ij}=0$, while the antiparallel-spin $\eta$ function may have a
nonzero derivative: see Appendix \ref{app:cusp}.  This accounts for
the differences in the behavior of the parallel- and antiparallel-spin
$\eta$ functions at small $r_{ij}$ which are visible in
Figs.~\ref{fig:heg_eta_AP} and~\ref{fig:heg_eta_P}.

The magnitude of our optimized three-electron Jastrow factor,
represented in Fig.~\ref{fig:heg_w}, increases monotonically with
$r_s$, and the maximum of $6(r_{ij}w_{ij})^2$ is at about
$r_{ij}/r_s=0.4$ for all densities.  This is in contrast with the
behavior of the three-electron Jastrow factor of Kwon \textit{et al.}\
(see Fig.~1 of Ref.\ \onlinecite{kwon_1998}), which changes sign at
$r_s=1$ (our parametrization is not allowed to do so) and breaks its
monotonicity with $r_s$ at $r_s=20$.  Kwon \textit{et al.}\ find that the maximum of the
plotted function is located at about $r_{ij}/r_s=1$.

\subsection{Lithium atom and dimer}

\subsubsection{AE lithium atom}

Our results for the ${}^1$S ground state of the AE lithium atom are
given in Table~\ref{table:ae_li}.  The SJ wave function gives a
reasonably good VMC energy.  Our backflow function consists of a
spin-pair-dependent e-e-n term with $N_{en}=N_{ee}=3$; this produces a
BF-VMC energy that is within statistical error bars of the exact value.
Note that the BF-VMC, SJ-DMC, and BF-DMC energies are within
statistical error bars of each other and are very close to the exact
value.  The excellent performance of the BF-VMC calculation is
particularly noteworthy.  The single-determinant nodal surface of
the ${}^1$S ground state of lithium is certainly extremely accurate and
may even be exact, although some contrary evidence has been
cited.\cite{bressanini_2002} It is therefore unlikely that backflow
could improve upon the SJ-DMC energy, and indeed it leaves it
essentially unchanged.

\begin{table}[!ht]
\begin{center}
\begin{tabular}{llrr@{.}lr@{.}lr@{.}l}
\hline \hline
Method & Wfn. & $N_p$ & \multicolumn{2}{c}{$E$ (a.u.)} &
\multicolumn{2}{c}{$\sigma^2$ (a.u.)} & \multicolumn{2}{c}{\% corr.~en.} \\
\hline
HF    &  - &   - & $-7$&$43273$   & \multicolumn{2}{c}{-} & $0$&$0$
\vspace{0.2cm}\\
Exact &  - &   - & $-7$&$47806$   & \multicolumn{2}{c}{-} & $100$&$0$
\vspace{0.2cm}\\
VMC   & SJ &   0 & $-7$&$47648(3)$ & $0$&$00385(2)$ & $96$&$52(8)$ \\
      & BF & 114 & $-7$&$47801(3)$ & $0$&$00241(1)$ & $99$&$89(6)$
\vspace{0.2cm}\\
DMC   & SJ &   0 & $-7$&$47803(8)$ & \multicolumn{2}{c}{-} &
 $99$&$9(2)$ \\
      & BF & 114 & $-7$&$47802(6)$ & \multicolumn{2}{c}{-} &
 $99$&$9(1)$ \\
\hline \hline
\end{tabular}
\caption{Slater-Jastrow and backflow results for the AE lithium atom.
The number of free backflow parameters, excluding cutoff lengths, is
$N_p$.  The Hartree-Fock (HF) and exact energies were taken from
Refs.\ \onlinecite{davidson_1991} and \onlinecite{chakravorty_1993}.
\label{table:ae_li}}
\end{center}
\end{table}

\subsubsection{AE lithium dimer}

\begin{figure}[!ht]
\begin{center}
\includegraphics[scale=0.5]{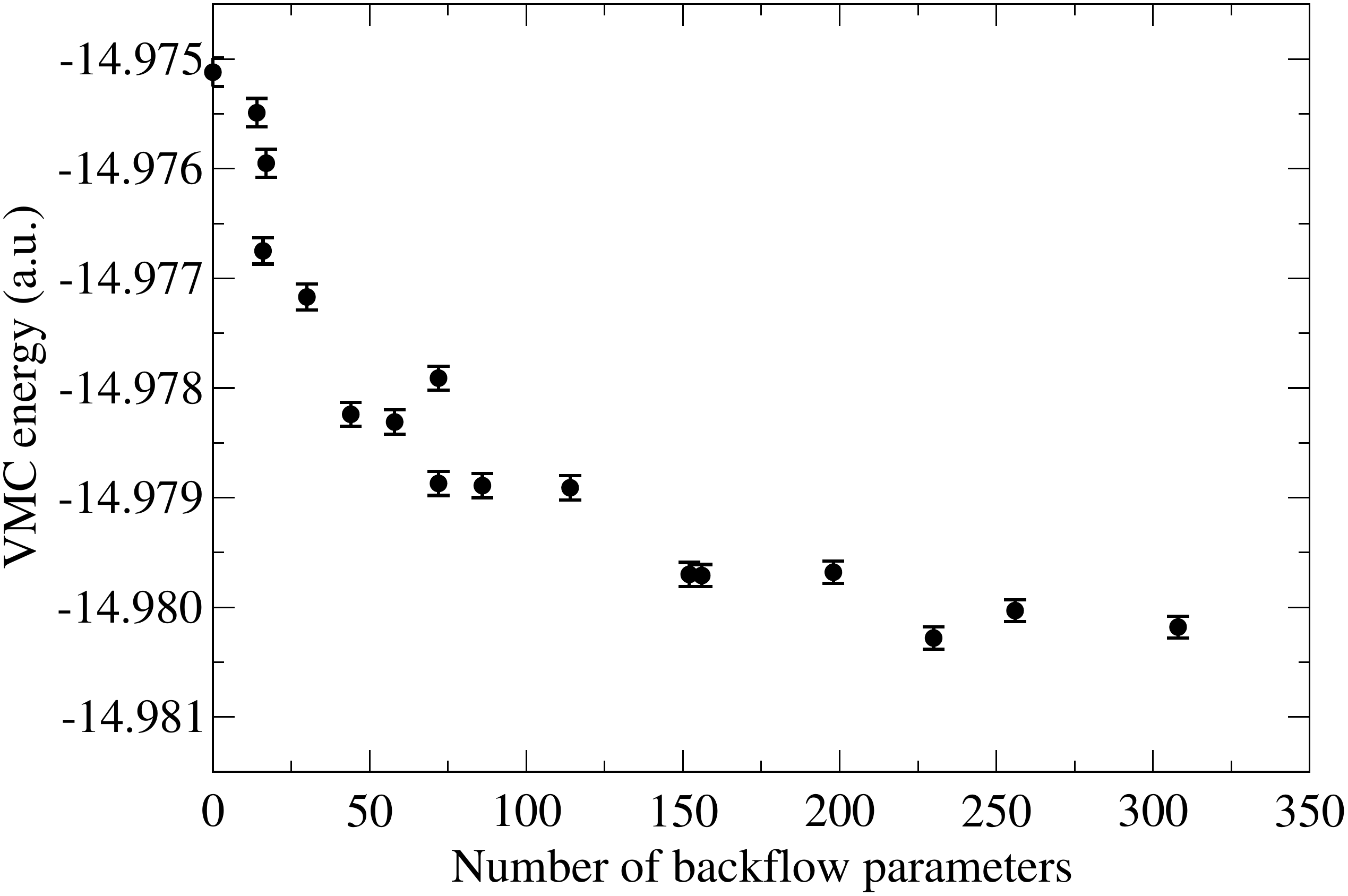}
\end{center}
\caption{The VMC energy of AE ${\rm Li}_2$ versus the total number of
parameters in the backflow functions.
\label{fig:ae_li2}}
\end{figure}

We studied the ground state of the AE ${\rm Li}_2$ dimer at the
experimental bond length of $5.051$~a.u.\cite{cade_1974} We tested
several different backflow functions, obtaining the results given in
Table~\ref{table:ae_li2}.  The use of homogeneous backflow retrieves
only an additional 0.7\% of the correlation energy.  A plot of the
VMC energy as a function of the number of parameters is displayed in
Fig.~\ref{fig:ae_li2}, which shows that the reduction in VMC energy is
very small beyond about 150 parameters.  Whereas backflow gave
99.89(6)\% of the correlation energy at the VMC level for the lithium
atom, for the dimer our best backflow transformation retrieves only
87.79(8)\%.  At the DMC level the improvement is small: using a SJ
wave function we obtain 96.2(3)\% of the correlation energy while
with the backflow wave function this improves slightly [to 97.1(3)\%].
Considerably better DMC results for ${\rm Li}_2$ have been obtained
using multideterminant (MD) wave functions.  Bressanini~\textit{et
al.}\cite{bressanini_2005b}\ obtained a DMC energy of $-14.9923(2)$
with one configuration state function (CSF), while their best result was $-14.9952(1)$ with 4 CSFs.

\begin{table}[!ht]
\begin{center}
\begin{tabular}{llccccrr@{.}lr@{.}lr@{.}l}
\hline \hline
Method & Wfn. & $N_\eta$ & $N_\mu$ & $N_{\rm en}$ & $N_{\rm ee}$ &
$N_p$ & \multicolumn{2}{c}{$E$ (a.u.)} & \multicolumn{2}{c}{$\sigma^2$ (a.u.)} &
\multicolumn{2}{c}{\% corr.~en.} \\
\hline
HF    & -& -& -& -& -&  -&$-14$&$871545$&\multicolumn{2}{c}{-}& $0$&$0$
\vspace{0.2cm}\\
Exact & -& -& -& -& -&  -&$-14$&$9954$  &\multicolumn{2}{c}{-}&$100$&$0$
\vspace{0.2cm}\\
VMC   &SJ& -& -& -& -&  0&$-14$&$9751(1)$& $0$&$0165(1)$  &$83$&$6(1)$ \\
      &BF& 0& 6& 0& 0& 14&$-14$&$9755(1)$& $0$&$01607(9)$ &$83$&$9(1)$ \\
      &BF& 8& 0& 0& 0& 17&$-14$&$9760(1)$& $0$&$01590(7)$ &$84$&$3(1)$ \\
      &BF& 0& 0& 2& 2& 16&$-14$&$9768(1)$& $0$&$01424(7)$ &$84$&$9(1)$ \\
      &BF& 0& 0& 2& 4& 44&$-14$&$9782(1)$& $0$&$01273(7)$ &$86$&$15(9)$\\
      &BF& 0& 0& 2& 6& 72&$-14$&$9789(1)$& $0$&$0125(1)$  &$86$&$65(9)$\\
      &BF& 0& 0& 3& 4&156&$-14$&$9797(1)$& $0$&$01102(5)$ &$87$&$33(8)$\\
      &BF& 0& 0& 4& 4&308&$-14$&$9802(1)$& $0$&$01030(6)$ &$87$&$71(8)$\\
      &BF& 0& 0& 4& 3&230&$-14$&$9803(1)$& $0$&$01038(4)$ &$87$&$79(8)$
\vspace{0.2cm}\\
DMC   &SJ& -& -& -& -&  0&$-14$&$9907(4)$&\multicolumn{2}{c}{-}&$96$&$2(3)$\\
      &BF& 0& 0& 4& 3&230&$-14$&$9918(4)$&\multicolumn{2}{c}{-}&$97$&$1(3)$\\
\hline \hline
\end{tabular}
\caption{Slater-Jastrow and backflow results for the AE ${\rm Li}_2$
molecule.  The different backflow forms have been put in order of decreasing
energy.  The number of free backflow parameters, excluding cutoff
lengths, is $N_p$.  The Hartree-Fock (HF) and exact energies were
taken from Ref.\ \onlinecite{filippi_1996}.
\label{table:ae_li2}}
\end{center}
\end{table}

The computed binding energies of ${\rm Li}_2$ are presented in
Table~\ref{table:ae_li2_bond}.  The BF-VMC, SJ-DMC, and BF-DMC energies
of the AE lithium atom are within error bars of the exact energy, and
therefore the error in the binding energy arises solely from the ${\rm Li}_2$
energy.  Backflow improves the VMC and DMC binding energies of ${\rm
Li}_2$ a little, but it is still somewhat short of the exact value.
The single-determinant nodal surface of ${\rm Li}_2$ is quite poor,
and backflow is not very effective at improving it.  Combining
MD wave functions with backflow might yield significant improvements
in this case.

\begin{table}[!ht]
\begin{center}
\begin{tabular}{llr@{.}l}
\hline \hline
Method & Wfn. & \multicolumn{2}{c}{$E_b$ (a.u.)} \\
\hline
HF    & - & $0$&$0061$
\vspace{0.2cm}\\
Exact & - & $0$&$0393$
\vspace{0.2cm}\\
VMC & SJ & $0$&$0221(1)$ \\
    & BF & $0$&$0243(1)$
\vspace{0.2cm}\\
DMC & SJ & $0$&$0346(4)$ \\
    & BF & $0$&$0358(4)$ \\
\hline \hline
\end{tabular}
\caption{Slater-Jastrow and backflow binding energies for the AE ${\rm
Li}_2$ molecule, computed using the best results from
Tables~\ref{table:ae_li} and \ref{table:ae_li2}.  The Hartree-Fock
(HF) and exact energies were taken from
Refs.\ \onlinecite{davidson_1991}, \onlinecite{chakravorty_1993}, and
\onlinecite{filippi_1996}, and references therein.
\label{table:ae_li2_bond}}
\end{center}
\end{table}


\subsection{Carbon atom, carbon dimer, and diamond}

\subsubsection{AE carbon atom}

\begin{figure}[!ht]
\begin{center}
\includegraphics[scale=0.5]{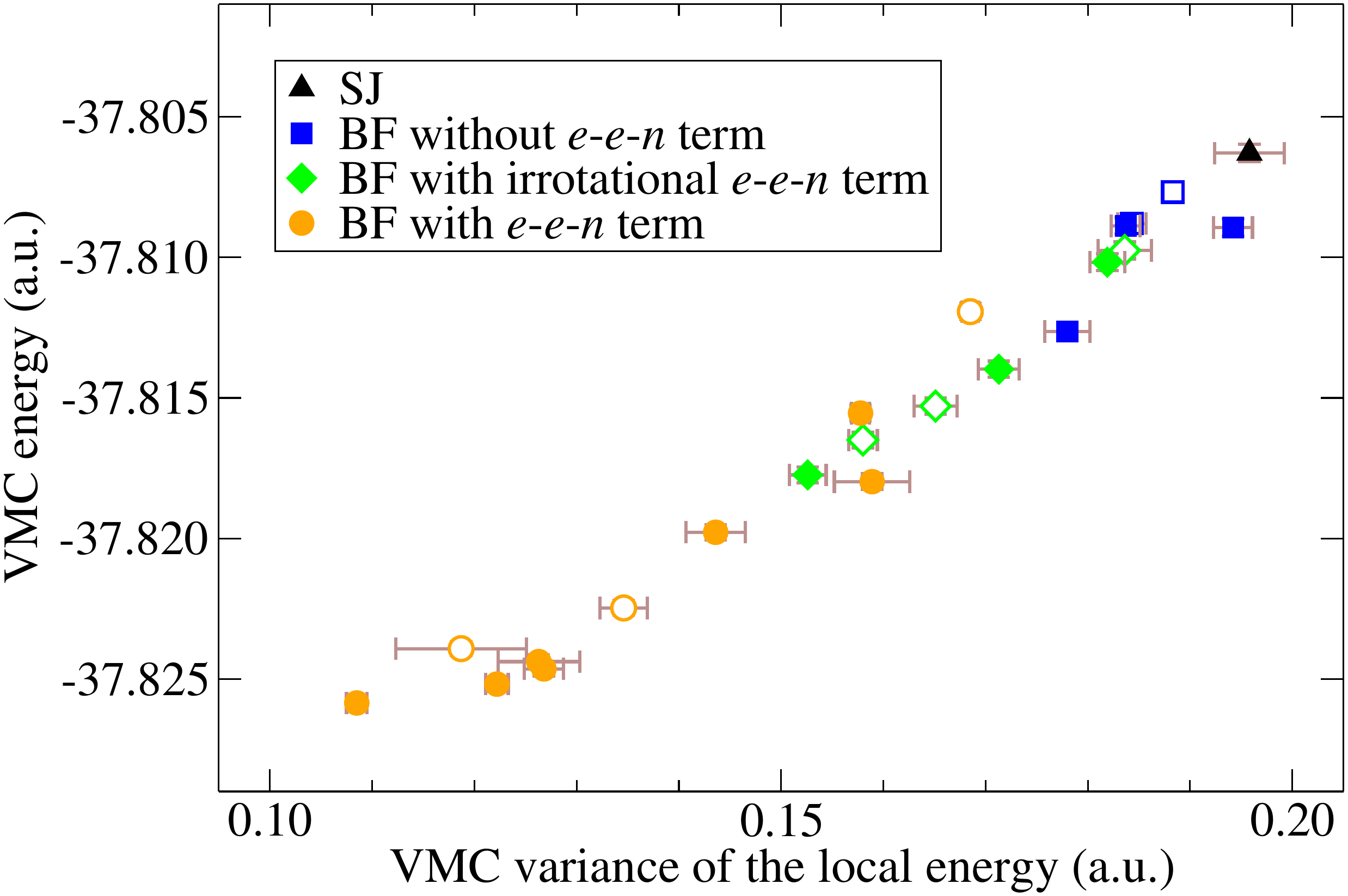}
\end{center}
\caption{(Color online) The VMC energy versus its variance for AE
carbon (see Table~\ref{table:ae_c}).  The open symbols denote that the
backflow parameters are independent of spin, while the filled symbols
denoted spin-dependent parameters.  The exact nonrelativistic energy
is $-37.8450$~a.u.\cite{davidson_1991,chakravorty_1993}
\label{fig:ae_c}}
\end{figure}

The ${}^3$P ground state of the AE carbon atom is a good example of a
system where single-determinant wave functions result in large
fixed-node errors: see Table~\ref{table:ae_c}.  In this case, we have
tested several combinations of terms, expansion orders, and constraints
to explore the possibilities of backflow transformations.  The VMC
data in Table~\ref{table:ae_c}, and additional data, are plotted in
Fig.~\ref{fig:ae_c}, where the performance of the different backflow
functions used can be compared conveniently.  Using only homogeneous
backflow (first BF-VMC results in Table~\ref{table:ae_c}) gives a very
small reduction in energy.  It seems that inhomogeneous systems
require inhomogeneous backflow to produce good wave functions, and the
e-e-n term is particularly successful in providing this.  To evaluate
the relative importance of the two e-e-n functions $\Phi_i^{jI}$ and
$\Theta_i^{jI}$, we performed calculations constraining the parameters
in one of them to be zero. The results are also given in
Table~\ref{table:ae_c}.  In this case $\Theta_i^{jI}$, which
contributes to the backflow displacement in the direction of ${\bf
r}_{iI}$, is slightly more important than $\Phi_i^{jI}$.  Applying
both terms gives better results than using only one of them, as we
expected.  We also tested the effect of constraining the backflow
displacement to be irrotational, which was suggested in
Ref.\ \onlinecite{holzmann_2003}.  The application of this
constraint, which is explained in Appendix \ref{app:nocurl},
approximately halves the number of parameters in the backflow
functions, but it gives very poor results for the carbon atom.

\begin{table}[!ht]
\begin{center}
\begin{tabular}{llcccccrr@{.}lr@{.}lr@{.}l}
\hline \hline
Method & Wfn. & $N_\eta$ & $N_\mu$ & $N_\Phi$ & S & I & $N_p$ &
\multicolumn{2}{c}{$E$ (a.u.)} &
\multicolumn{2}{c}{$\sigma^2$ (a.u.)} & \multicolumn{2}{c}{\% corr.~en.} \\
\hline
HF    & -& -& -& -& -& -&  -&$-37$&$688619$&\multicolumn{2}{c}{-}& $0$&$0$
\vspace{0.2cm}\\
Exact & -& -& -& -& -& -&  -&$-37$&$8450$  &\multicolumn{2}{c}{-}&$100$&$0$
\vspace{0.2cm}\\
VMC   &SJ& -& -& -& -& -&  0& $-37$&$8064(3)$ & $0$&$193(2)$  & $75$&$3(2)$ \\
      &BF& 8& 0& 0& T& -& 17& $-37$&$8089(3)$ & $0$&$194(2)$  & $76$&$9(2)$ \\
      &BF& 0& 6& 0& T& -& 10& $-37$&$8089(3)$ & $0$&$184(1)$  & $76$&$9(2)$ \\
      &BF& 0& 0& 2& F& F& 10& $-37$&$8119(3)$ & $0$&$1685(8)$ & $78$&$8(2)$ \\
      &BF& 8& 6& 0& T& -& 27& $-37$&$8126(3)$ & $0$&$178(2)$  & $79$&$3(2)$ \\
      &BF& 0& 0& 4& T& T& 35& $-37$&$8140(3)$ & $0$&$171(2)$  & $80$&$2(2)$ \\
&BF${}^\dag$&0&0&3& T& F& 56& $-37$&$8155(3)$ & $0$&$1578(9)$ & $81$&$1(2)$ \\
      &BF& 0& 0& 5& T& T&121& $-37$&$8177(3)$ & $0$&$153(2)$  & $82$&$5(2)$ \\
      &BF& 0& 0& 2& T& F& 16& $-37$&$8180(3)$ & $0$&$159(4)$  & $82$&$7(2)$ \\
&BF${}^\ddag$&0&0&3&T& F& 58& $-37$&$8198(3)$ & $0$&$144(2)$  & $83$&$7(2)$ \\
      &BF& 0& 0& 3& F& F& 60& $-37$&$8225(3)$ & $0$&$135(2)$  & $85$&$5(2)$ \\
      &BF& 0& 0& 4& F& F&158& $-37$&$8239(3)$ & $0$&$119(6)$  & $86$&$5(2)$ \\
      &BF& 0& 0& 3& T& F&114& $-37$&$8246(3)$ & $0$&$127(2)$  & $87$&$0(2)$ \\
      &BF& 0& 6& 3& T& F&124& $-37$&$8252(3)$ & $0$&$122(1)$  & $87$&$3(2)$ \\
      &BF& 0& 0& 4& T& F&308& $-37$&$8259(3)$ & $0$&$109(1)$  & $87$&$8(2)$
\vspace{0.2cm}\\
DMC   &SJ& -& -& -& -& -&  0& $-37$&$8297(2)$ & \multicolumn{2}{c}{-} &
 $90$&$2(1)$ \\
      &BF& 0& 6& 3& T& F&124& $-37$&$8324(1)$ & \multicolumn{2}{c}{-} &
 $92$&$0(1)$ \\
\hline \hline
\end{tabular}
\caption{Slater-Jastrow and backflow results for the AE carbon atom.
The different backflow forms have been put in order of decreasing energy.
Key: $N_\Phi \equiv N_{\rm en}=N_{\rm ee}$; S indicates whether the
parameters are spin and spin-pair dependent (T) or not (F); I
indicates whether the constraints for irrotational backflow have been
applied (T) or not (F); $N_p$ is the number of free backflow
parameters, excluding cutoff lengths.  Where the {\dag} symbol is
used, we constrained $\theta_{klm,I}=0$; where the {\ddag} symbol
appears, we constrained $\varphi_{klm,I}=0$.  The Hartree-Fock (HF)
and exact energies were taken from
Refs.\ \onlinecite{davidson_1991} and \onlinecite{chakravorty_1993}.
\label{table:ae_c}}
\end{center}
\end{table}

The most satisfactory backflow forms reduce the difference between the
VMC and exact energies by a factor of about 2.  The further energy
reduction from using DMC is quite small, and our BF-DMC calculation
gave an energy of $-37.8324(1)$~a.u., which corresponds to 92.0(1)\%
of the total correlation energy.  This suggests that, although
backflow improves significantly upon the single-determinant nodal
surface of the carbon atom, it misses some important features of the
exact nodal surface.  It is well-known that the single-determinant
nodal surface of the carbon atom can be substantially improved by
using MD trial wave functions.  Barnett \textit{et
al.}\cite{barnett_2000}\ used an MD trial wave function consisting of
14 CSFs, and obtained a DMC energy of
$-37.8420(3)$~a.u., which corresponds to 98.1(2)\% of the correlation
energy.  Glauser \textit{et al.}\cite{glauser_1992}\ showed that the
configuration space of a single-determinant of HF orbitals for the
${}^3$P ground state carbon atom is divided into four nodal
pockets,\footnote{Two configurations are in the same nodal pocket if
there exists a continuous path between the two along which the wave
function does not change sign and is not equal to zero.  Nodal pockets
are bounded by nodal surfaces, which determine the shape and number of
the former.} but more accurate wave functions indicate that the exact
wave function has two nodal pockets.  It appears that backflow
transformations are unable to correct this defect in the
single-determinant nodal surface.

\subsubsection{PP carbon atom}

We have also studied how backflow performs in systems where PPs are
used.  Our results for a PP carbon atom are given in Table~\ref{table:pp_c}.  The reduction in the
VMC energy obtained with backflow is much smaller than for the AE
carbon atom, but the corresponding energy reduction within DMC of
$0.0039(1)$~a.u.\ is somewhat larger than the AE one of
$0.0027(2)$~a.u.  A peculiarity of this case is that the reduction in
the DMC energy resulting from the use of backflow is $71\%$ of the reduction in the
VMC energy, which is the largest such percentage amongst the
calculations described here.

\begin{table}[!ht]
\begin{center}
\begin{tabular}{llrr@{.}lr@{.}lr@{\@}l}
\hline \hline
Method & Wfn. & $N_p$ & \multicolumn{2}{c}{$E$ (a.u.)} &
\multicolumn{2}{c}{$\sigma^2$ (a.u.)} & \multicolumn{2}{c}{\% corr.~en.} \\
\hline
HF    &  - &   - & $-5$&$31663$  & \multicolumn{2}{c}{-} & $0$&$.0$
\vspace{0.2cm}\\
Exact &  - &   - & $-5$&$420(4)$  & \multicolumn{2}{c}{-} & $100$&$(8)$
\vspace{0.2cm}\\
VMC   & SJ &   0 & $-5$&$4007(1)$ & $0$&$0582(4)$ & $81$&$(3)$ \\
      & BF & 218 & $-5$&$4061(1)$ & $0$&$0502(6)$ & $87$&$(3)$
\vspace{0.2cm}\\
DMC   & SJ &   0 & $-5$&$40886(7)$ & \multicolumn{2}{c}{-} &
 $89$&$(4)$ \\
      & BF & 218 & $-5$&$41273(9)$ & \multicolumn{2}{c}{-} &
 $93$&$(4)$ \\
\hline \hline
\end{tabular}
\caption{Slater-Jastrow and backflow results for the PP carbon atom.
The number of free backflow parameters, excluding cutoff lengths, is
$N_p$.  The exact energy was taken from Ref.\ \onlinecite{dolg_1996}.
\label{table:pp_c}}
\end{center}
\end{table}


\subsubsection{PP carbon dimer}

For the PP carbon dimer we used the experimental bond length of
$2.3622$~a.u.,\cite{cade_1974} obtaining the results given in
Table~\ref{table:pp_c2}.
The carbon dimer is another example of a system in which MD effects
are known to be substantial.  Backflow results in larger energy
reductions per atom than for the isolated atom at both the VMC and DMC
levels.  The computed binding energies of ${\rm C}_2$ are presented in
Table~\ref{table:pp_c2_bond}.  The use of backflow slightly improves
the binding energy of the dimer.

\begin{table}[!ht]
\begin{center}
\begin{tabular}{llrr@{.}lr@{.}lr@{\@}l}
\hline \hline
Method & Wfn. & $N_p$ & \multicolumn{2}{c}{$E$ (a.u.)} &
\multicolumn{2}{c}{$\sigma^2$ (a.u.)} & \multicolumn{2}{c}{\% corr.~en.} \\
\hline
HF    &  - &   - & $-10$&$652399$  & \multicolumn{2}{c}{-} & $0$&$.0$
\vspace{0.2cm}\\
Exact &  - &   - & $-11$&$055(4)$  & \multicolumn{2}{c}{-} & $100$&$(2)$
\vspace{0.2cm}\\
VMC   & SJ &   0 & $-10$&$9870(3)$ & $0$&$168(1)$ & $83$&$.1(9)$ \\
      & BF & 214 & $-11$&$0173(2)$ & $0$&$156(2)$ & $91$&$(1)$
\vspace{0.2cm}\\
DMC   & SJ &   0 & $-11$&$0237(4)$ & \multicolumn{2}{c}{-} &
 $92$&$(1)$ \\
      & BF & 214 & $-11$&$0348(6)$ & \multicolumn{2}{c}{-} &
 $95$&$(1)$ \\
\hline \hline
\end{tabular}
\caption{Slater-Jastrow and backflow results for the PP ${\rm C}_2$
molecule.  The number of free backflow parameters, excluding cutoff
lengths, is $N_p$.  The exact energy was taken from
Ref.\ \onlinecite{dolg_1996}.
\label{table:pp_c2}}
\end{center}
\end{table}

\begin{table}[!ht]
\begin{center}
\begin{tabular}{llr@{.}l}
\hline \hline
Method & Wfn. & \multicolumn{2}{c}{$E_b$ (a.u.)} \\
\hline
HF    &  - & $0$&$02896$
\vspace{0.2cm}\\
Exact &  - & $0$&$233(5)$
\vspace{0.2cm}\\
VMC   & SJ & $0$&$1856(3)$ \\
      & BF & $0$&$2051(2)$
\vspace{0.2cm}\\
DMC   & SJ & $0$&$2060(4)$ \\
      & BF & $0$&$2093(6)$ \\
\hline \hline
\end{tabular}
\caption{Slater-Jastrow and backflow binding energies for the PP ${\rm
C}_2$ molecule, computed using the results from
Tables~\ref{table:pp_c} and \ref{table:pp_c2}.  The Hartree-Fock (HF)
and exact energies were taken from Refs.\ \onlinecite{davidson_1991},
\onlinecite{chakravorty_1993}, and \onlinecite{filippi_1996}, and the
references therein.  The data in Ref.~\onlinecite{dolg_1996} can be
used to estimate an approximate binding energy of $0.215(6)$~a.u..
\label{table:pp_c2_bond}}
\end{center}
\end{table}


\subsubsection{PP diamond}

We have also studied PP carbon diamond with the experimental cubic
lattice constant of $6.741$~a.u.,\cite{sato_2002} representing
the solid by a $2\times 2\times 2$
supercell containing 16 atoms subject to periodic boundary conditions.
Diamond is an insulator with a large band gap, and therefore we expect
the single-determinant nodal surface to be quite accurate.
We parametrized our backflow function using $N_\eta=8$,
$N_\mu=8$, and $N_{\rm ee}=N_{\rm en}=2$, allowing all parameters to be
spin and spin-pair dependent.  The cutoff lengths were optimized, and
they went to the maximum allowed values.  The results in
Table~\ref{table:c_diam} show that backflow gives a substantial
reduction in the VMC energy of $0.0131(2)$~a.u.\ per atom
[$0.356(5)$~eV per atom], which is accompanied by a reduction in the
variance by a factor of nearly two.  The reduction in the VMC energy
of diamond from using backflow is somewhat smaller than that obtained
in the dimer [$0.411(5)$~eV per atom], and substantially larger than
that in the atom [$0.147(3)$~eV per atom].  This may arise from the
fact that the backflow functions in diamond are quite long ranged and
cover several atoms. Backflow reduces the DMC energy of diamond by
$0.0035(2)$~a.u.\ [$0.095(5)$~eV per atom] per atom, which is a little
less than in the dimer [$0.15(1)$~eV per atom] and atom [$0.106(3)$~eV
per atom].


\begin{table}[!ht]
\begin{center}
\begin{tabular}{llrr@{.}lr@{.}l}
\hline \hline
Method & Wfn. & $N_p$ & \multicolumn{2}{c}{$E$ (a.u./prim.~cell)} &
\multicolumn{2}{c}{$\sigma^2$ (a.u.)} \\
\hline
DFT-PBE &  - &   - & $-11$&$368208$   & \multicolumn{2}{c}{-}
\vspace{0.2cm}\\
VMC     & SJ &   0 & $-11$&$3708(2)$  & $1$&$51(8)$ \\
        & BF &  96 & $-11$&$3970(3)$  & $0$&$897(8)$
\vspace{0.2cm}\\
DMC     & SJ &   0 & $-11$&$40717(8)$ & \multicolumn{2}{c}{-} \\
        & BF &  96 & $-11$&$4141(3)$  & \multicolumn{2}{c}{-} \\
\hline \hline
\end{tabular}
\caption{Slater-Jastrow and backflow energies per primitive cell for
PP carbon diamond using a face-centered cubic cell containing 16
atoms.  The number of free backflow parameters, excluding cutoff
lengths, is $N_p$.
\label{table:c_diam}}
\end{center}
\end{table}

We do not discuss the cohesive energy of the diamond crystal, as we
would need to account for finite-size effects to be able to compare
with experimental data.  Within VMC, the energy gain per atom from
using backflow is larger in the solid than in the atom, and hence the
cohesive energy is substantially reduced.  Within DMC, both the solid
and the atom present a similar energy gain per atom, and the cohesive
energy is not changed significantly.


\section{Discussion}
\label{sec:discuss}


\subsection{Electron-by-electron and configuration-by-configuration algorithms}
\label{sec:discuss_ebea-cbca}

The additional complexity of BF wave functions compared with SJ ones
leads to greater computational expense in QMC calculations.  One of
the most costly operations in QMC calculations is the evaluation of
the orbitals and their first two derivatives at points in the
configuration space.  The evaluation of the collective coordinates
involves some extra cost.  Furthermore, while QMC calculations with SJ
wave functions
require only the value, gradient, and Laplacian of each orbital $\phi$,
calculations with BF wave functions also require cross derivatives such
as $\partial^2 \phi/ \partial x \partial y$, as explained in
Ref.\ \onlinecite{kwon_1993}.  However, the most important complicating factor
arising from backflow transformations is that they make each orbital in the
Slater determinants depend on the coordinates of every particle.  In
standard QMC algorithms with SJ wave functions one moves the electrons
sequentially in what we call the electron-by-electron algorithm
(EBEA)\@.  Fast update algorithms are used in the EBEA to replace
altered rows in the Slater determinants efficiently and the
accept/reject step is performed on each particle separately.  However,
in BF calculations each collective coordinate depends on every
electron position and therefore the fast update algorithms used in
the EBEA are no longer appropriate, so that one must recalculate the
determinants at each step using LU decomposition.  Nevertheless,
the implementation of the EBEA for backflow wave functions can take
advantage of other optimizations to make the algorithm more
efficient, such as buffering the separate contributions to the
collective coordinates, which we have exploited as far as possible.

In previous fermion backflow
calculations\cite{kwon_1993,kwon_1998,holzmann_2003} the electrons
have all been moved together and a single accept/reject step has
been performed, in what we call the
configuration-by-configuration algorithm (CBCA)\@.  We have compared the
efficiency of the EBEA and CBCA\@.  The relative efficiency of the EBEA
and CBCA depends on the computational costs of moving the electrons
and the correlation time of the local energies, which is proportional
to the number of moves of all the electrons required before the local
energies are uncorrelated.  Let A and B be two calculations for the
same system, identical except for the use of different sampling
algorithms.  We define the \textit{relative efficiency} $\gamma$ of A
and B as
\begin{equation}
\label{eq:rel_eff}
\gamma({\rm A},{\rm B}) = \frac{t_{\rm A}\sigma_{\rm A}^2} {t_{\rm
    B}\sigma_{\rm B}^2} \;,
\end{equation}
where $t$ is the CPU time and $\sigma$ is the standard error in the
mean energy.\cite{flyvbjerg_1989} $\gamma$ represents the ratio of the
time required to achieve a fixed error in the mean energy in
calculation A to that required in calculation B, and is hence appropriate for
comparing the efficiency of the two algorithms.

In Table~\ref{table:ebea_cbca} we report results for the systems
studied in this paper.  For each system, the EBEA and CBCA time steps
were chosen so that the same proportion of proposed moves were
accepted: in VMC the target acceptance ratio was 50\%, which
corresponds to fairly efficient sampling, and in DMC it was around
99.5\%.  The correlation times for the CBCA are considerably longer
than for the EBEA\@.  The ratio of the correlation time of the CBCA to
that of the EBEA (the ``correlation time ratio'' or CTR) appears to
increase roughly linearly with the number of atoms (for example,
compare the AE Li atom and ${\rm Li}_2$ molecule, and the PP C atom,
${\rm C}_2$ molecule, and diamond), or with the number of electrons.
$\gamma({\rm CBCA},{\rm EBEA})$ is larger than unity in all cases
except the BF-VMC calculation of lithium atom, so the EBEA is
generally found to be more
efficient than the CBCA\@.  $\gamma({\rm CBCA},{\rm EBEA})$ is larger
for SJ wave functions than for BF ones because for SJ wave functions
and the EBEA one uses fast update algorithms.

\begin{table}[!ht]
\begin{center}
\begin{tabular}{lrrcr@{.}lr@{.}l}
\hline \hline
System & $N_e$ & Wfn & CTR. &\multicolumn{2}{c}{$\gamma_{\rm VMC}$}&
\multicolumn{2}{c}{$\gamma_{\rm DMC}$}\\
\hline
HEG ($r_s=1.0$)          & 54 & SJ & 70 &  44&0  & 7&5 \\
                         &    & BF &    &  34&0  & 1&2
\vspace{0.2cm}\\
AE Li atom               &  3 & SJ &  9 &   4&6  & 1&8 \\
                         &    & BF &  1 &   0&43 & 1&9 \\
AE ${\rm Li}_2$ molecule &  6 & SJ & 15 &  15&4  & 5&0 \\
                         &    & BF &    &   9&9  & 3&8
\vspace{0.2cm}\\
AE C atom                &  6 & SJ &  9 &   5&1  & 7&9 \\
                         &    & BF &  3 &   2&4  & 3&7 \\
PP C atom                &  4 & SJ &  3 &   3&3  & 3&7 \\
                         &    & BF &  2 &   2&3  & 1&3 \\
PP ${\rm C}_2$ molecule  &  8 & SJ & 10 &   6&9  & 3&5 \\
                         &    & BF &    &   6&9  & 1&5 \\
PP C diamond             & 64 & SJ & 80 &  99&0  & \multicolumn{2}{c}{-} \\
($2\times 2\times 2$)    &    & BF &    &  39&0  & \multicolumn{2}{c}{-} \\
\hline \hline
\end{tabular}
\caption{Comparison of the EBEA and CBCA for SJ and BF wave functions.
Key: $N_e$ is the number of electrons; ``CTR'' is the ratio of the
correlation time in the CBCA to that in the EBEA; $\gamma$ is
$\gamma$(CBCA,EBEA), as defined in Eq.~(\ref{eq:rel_eff}).  Where a
separate ``CTR'' for the BF wave function has not been reported, it is
because it was found to equal that of the SJ wave function.
\label{table:ebea_cbca}}
\end{center}
\end{table}

Apart from the tests reported in this section, all the VMC and DMC
results reported in this paper have been obtained using the EBEA\@.


\subsection{Computational expense of backflow calculations}
\label{sec:discuss_cost}

We now investigate the relative cost of BF and SJ calculations.
The additional computational expense of each step in a BF calculation is
offset by the fact that
BF wave functions are generally more accurate than SJ ones, so that the
variance of the energy is smaller, and consequently the number of
statistically independent local energies required to achieve a given
error bar in the mean energy is also smaller.

Let A and B be two calculations
for the same system, identical except for the use of different
wave functions.  We define the \textit{time ratio} as
$\tau({\rm A},{\rm B})=t_{\rm A}/t_{\rm B}$, the \textit{squared-error
ratio} as $\epsilon({\rm A},{\rm B})=\sigma_{\rm A}^2/\sigma_{\rm B}^2$,
and the \textit{relative efficiency} as
$\gamma({\rm A},{\rm B})=\tau({\rm A},{\rm B})\epsilon({\rm A},{\rm B})$,
where $t$ is the CPU time and $\sigma$ is the standard error in the mean energy.
$\tau({\rm A},{\rm B})$ represents the relative expense per move of
calculation A with respect to calculation B or, equivalently, the
relative expense of generating a fixed number of configurations.  The
latter is relevant to the wave-function optimization procedure,
as the number of configurations used should, if anything, increase
with the number of parameters in the wave function.
$\gamma({\rm A},{\rm B})$ measures
the relative ability of calculation A to produce a total energy to
a desired degree of certainty with respect to B.

In Table~\ref{table:timing} we compile the BF-to-SJ ratios $\tau$,
$\epsilon$, and $\gamma$ for each calculation.  For the HEG at
$r_s=20$ we report BF3-to-SJ3 ratios instead.  The performance of
backflow in the HEG is impressive: backflow not only improves the
energies but also makes the calculations less costly!

The lithium atom is another example of improved efficiency.  According
to Table~\ref{table:ae_li}, the SJ-DMC and BF-DMC energies are equal,
so it does not seem advantageous to use backflow at all in this
system.  However, due to the BF-VMC energy being so close to the
BF-DMC value, the variance of the BF-DMC run is enormously lowered and
the CPU time is reduced to $25\%$ of the time taken by the SJ-DMC
run.\footnote{The variance of the local energies encountered during a
DMC calculation is approximately proportional to $E_{\rm VMC}-E_{\rm
DMC}$.\cite{ceperley_1986,ma_2005b}}

In all cases except PP diamond, $\gamma$ is less than $3$ in VMC,
and $6$ in DMC\@.  However, the crystalline PP calculations become significantly
more expensive when backflow is used.  A great part of this increase
is due to the computation of the nonlocal PP energy, which involves
several evaluations of the wave function ($12$, in this case) for each
electron and each ion, every time the local energy is computed.

\begin{table}[!ht]
\begin{center}
\begin{tabular}{lrcr@{.}lr@{.}lr@{.}l}
\hline \hline
System & $N_e$ & Method & \multicolumn{2}{c}{$\tau$} &
\multicolumn{2}{c}{$\epsilon$} & \multicolumn{2}{c}{$\gamma$} \\
\hline
HEG ($r_s=1.0$)          & 54 & VMC &  2&9 & 0&18 &  0&52\\
                         &    & DMC &  4&9 & 0&15 &  0&75\\
HEG ($r_s=20.0$)         & 54 & VMC &  1&4 & 0&48 &  0&67\\
                         &    & DMC &  2&0 & 0&15 &  0&28
\vspace{0.2cm}\\
AE Li atom               &  3 & VMC &  2&3 & 0&52 &  1&2 \\
                         &    & DMC &  4&3 & 0&06 &  0&25\\
AE ${\rm Li}_2$ molecule &  6 & VMC &  3&9 & 0&71 &  2&8 \\
                         &    & DMC &  8&3 & 0&71 &  5&9
\vspace{0.2cm}\\
AE C atom                &  6 & VMC &  3&5 & 0&69 &  2&4 \\
                         &    & DMC &  5&9 & 0&41 &  2&4 \\
PP C atom                &  4 & VMC &  3&1 & 0&79 &  2&4 \\
                         &    & DMC &  2&8 & 0&73 &  2&1 \\
PP ${\rm C}_2$ molecule  &  8 & VMC &  3&9 & 0&65 &  2&5 \\
                         &    & DMC &  3&3 & 0&47 &  1&6 \\
PP C diamond             & 64 & VMC & 27&0 & 0&31 &  8&3 \\
($2\times 2\times 2$)    &    & DMC & \multicolumn{2}{c}{-} &
                          \multicolumn{2}{c}{-} & \multicolumn{2}{c}{-} \\
\hline \hline
\end{tabular}
\caption{Data from the timing tests performed on different systems.
Key: $N_e$ is the number of electrons; $\tau$, $\epsilon$, and $\gamma$
are $\tau$(BF,SJ), $\epsilon$(BF,SJ), and $\gamma$(BF,SJ), as defined
in the text.
\label{table:timing}}
\end{center}
\end{table}


\subsection{Backflow and nodes}
\label{sec:discuss_nodes}

HF nodes have been compared with either exact or very accurate nodes in a
number of
studies.\cite{glauser_1992,bressanini_2002,bressanini_2005a,bressanini_2005b,bajdich_2005}
It has been found that the HF wave function often has too many nodal
pockets for the ground states of atoms with four or more electrons.
It is conceivable that coordinate transformations could modify the
number of nodal pockets of a wave function.  However, we believe this
to be unlikely for the backflow transformation presented in this paper,
because this would require the backflow displacement field to be
discontinuous at very specific configurations, or exhibit other
unusual features.  The development of a general backflow
transformation with the \textit{appropriate} discontinuities to
correct HF nodes, which we have not attempted, seems likely to be a
tremendously difficult task.

\begin{figure}[!ht]
\begin{center}
\subfigure[~Nodes in a 2D HEG at $r_s=0.5$.]{
\includegraphics[height=80mm,angle=270]{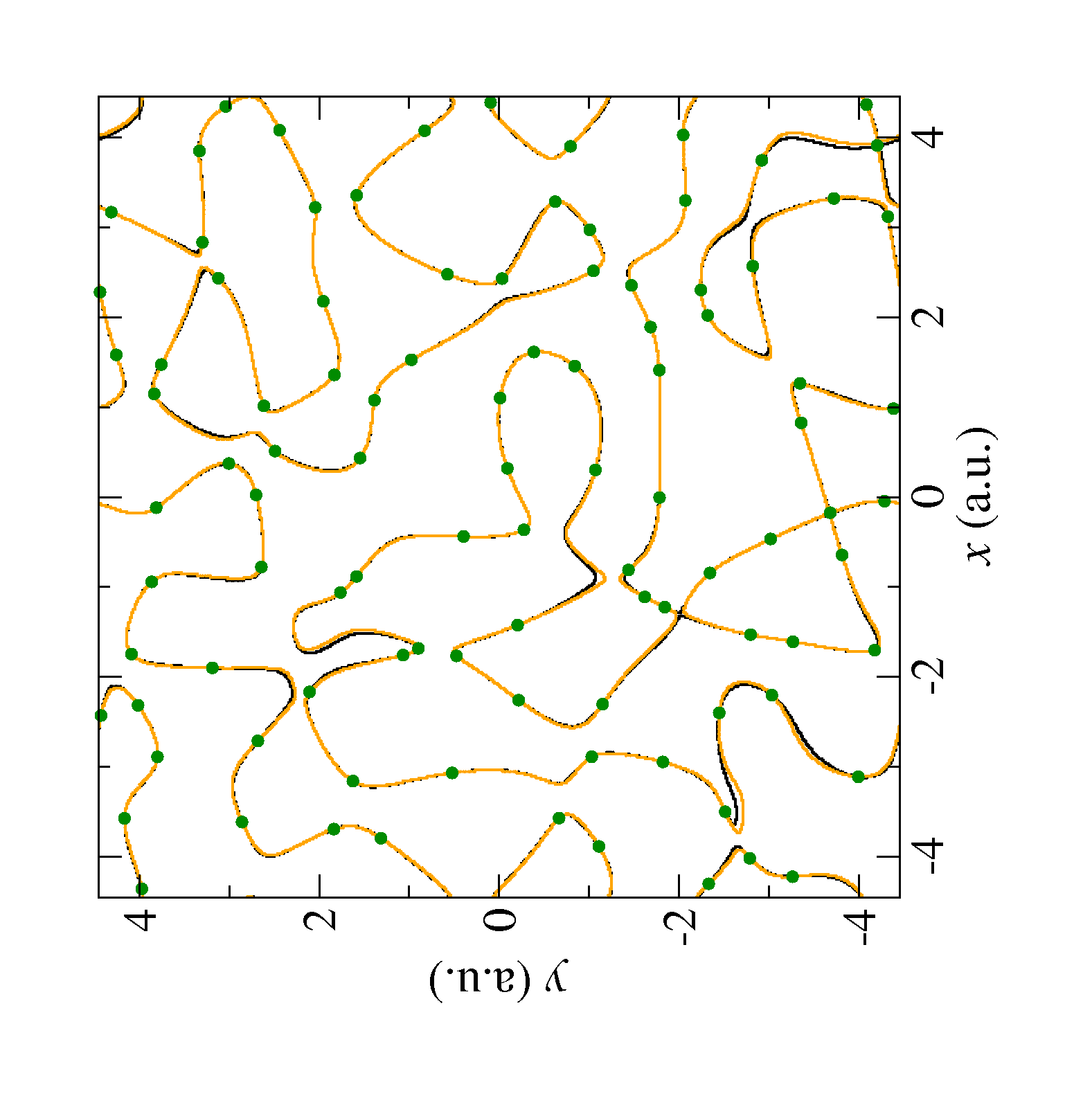}}
\hspace{0.5cm}
\subfigure[~Nodes in a 2D HEG at $r_s=10$.]
{\includegraphics[height=80mm,angle=270]{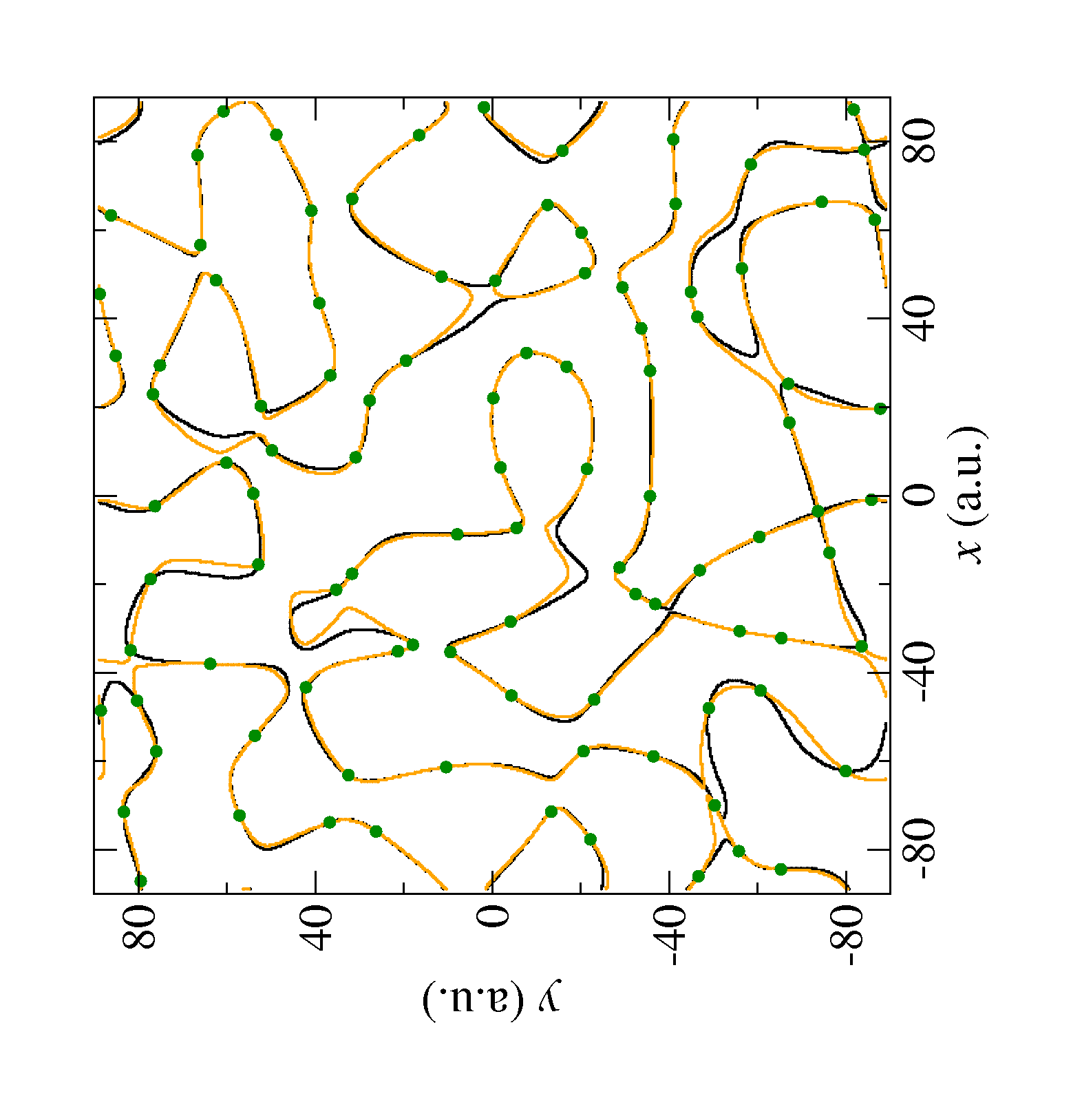}}
\end{center}
\caption{(Color online) Nodes encountered when moving one of the
electrons of a two-dimensional HEG of 101 like-spin electrons at two
different densities.  The HF and BF nodes are in black and orange,
respectively.  The green circles indicate the
positions of the remaining 100 electrons, at which the nodes are
required to remain fixed.  The backflow wave functions were obtained by
variance minimization; the energy reductions from SJ-VMC to BF-VMC
at $r_s=0.5$ and $10$ were $0.0007(1)$~a.u./electron and
$0.00005(1)$~a.u./electron, respectively.
\label{fig:2Dheg_nodes}}
\end{figure}

We now illustrate graphically how our backflow transformations changes
nodal surfaces.  Note that the figures described below are single
projections of high-dimensional nodal surfaces, from which almost no
useful conclusions regarding the full topology of the nodes can be
extracted.  Two-dimensional projections of the HF and BF nodes for a
two-dimensional HEG are depicted in Fig.~\ref{fig:2Dheg_nodes} at two
different densities.  The effect of backflow on the nodes is much more
pronounced for the low-density HEG\@.  For an unpolarized system, the
nodal changes should be larger than those seen in Fig.~\ref{fig:2Dheg_nodes}
at all densities.  Some regions of
these plots suggest that the displacement of the nodes due to backflow
is largest at points where the curvature of the nodal surface is large,
away from electron-electron coalescences.  There are a number of
avoided crossings in Fig.~\ref{fig:2Dheg_nodes} (3 at $r_s=0.5$, and 6
at $r_s=10$) whose connectivity (in the projection) is modified by
backflow.

\begin{figure}[!ht]
\begin{center}
\subfigure[~HF nodes of a carbon atom.]
{\includegraphics[height=80mm]{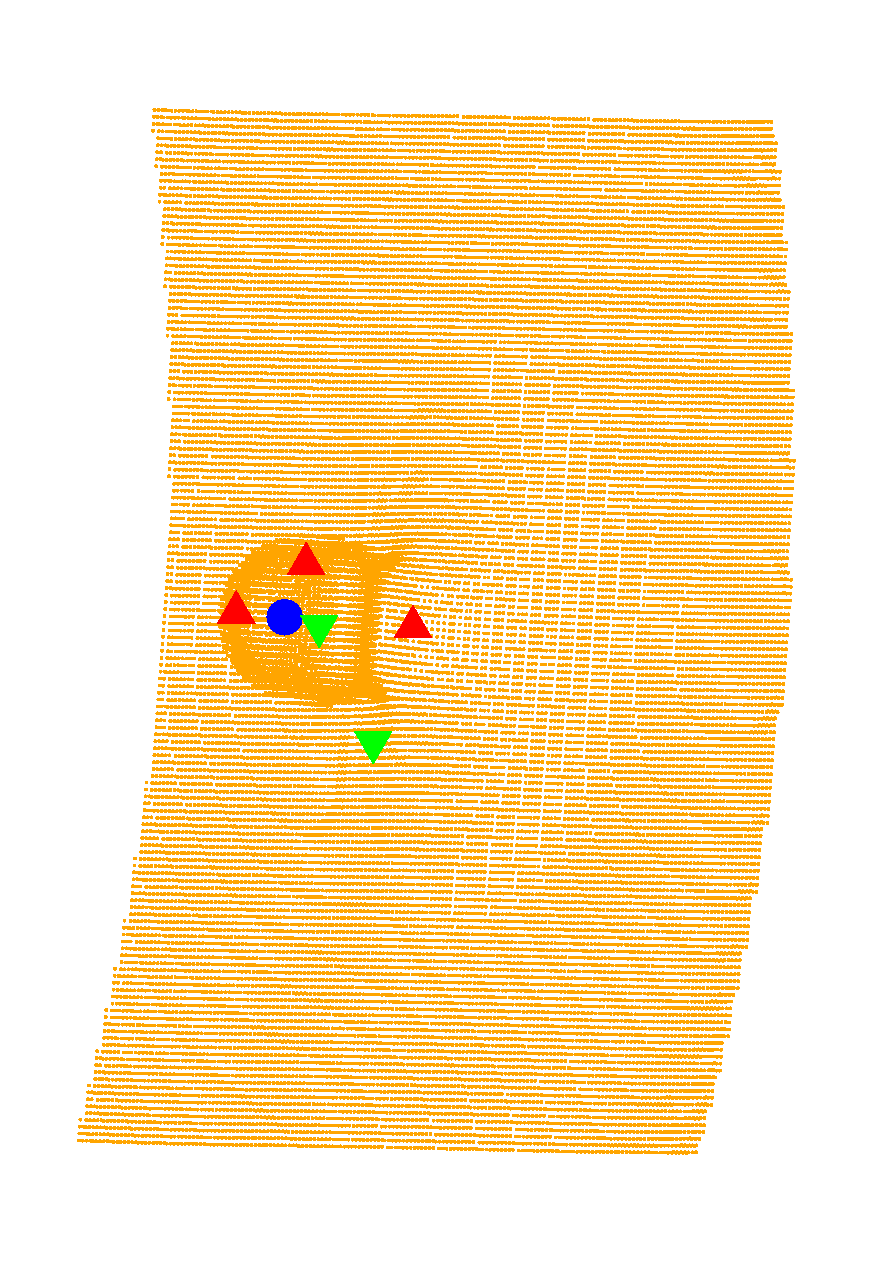}}
\hspace{0.5cm}
\subfigure[~BF nodes of a carbon atom.]
{\includegraphics[height=80mm]{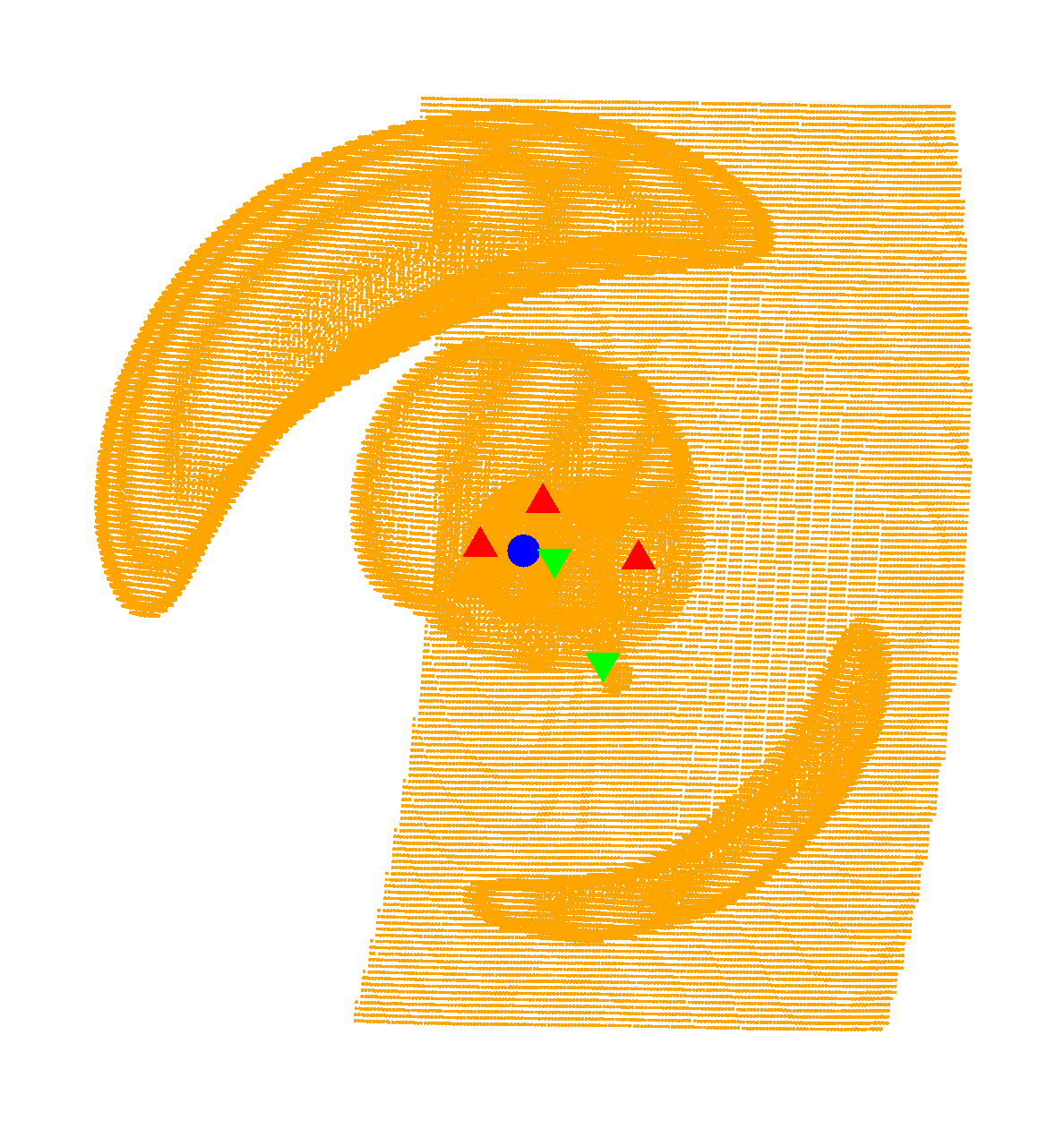}}
\end{center}
\caption{(Color online) HF and BF nodes encountered
when moving one of the (majority spin) up-spin electrons of an AE
carbon atom.  The blue circle corresponds to the position of the
nucleus, the red upward-pointing triangles indicate the positions of
the remaining up-spin electrons and the green downward-pointing
triangles indicate the positions of the down-spin ones.  The HF node
consists of a (seemingly) infinite sheet with a bubble attached to it,
which contains the nucleus.  Backflow slightly modifies this node, and
adds three large lobes (detached from one another; all intersect the
HF node) and a small bubble next to a down-spin electron.
\label{fig:carbon_nodes}}
\end{figure}

Three-dimensional projections of the HF and BF nodes of the AE carbon
atom are compared in Fig.~\ref{fig:carbon_nodes}.  The nodes are
substantially modified by the introduction of backflow.  New nodal
regions appear in this projection because the electron being moved
``pushes'' the other electrons (via the backflow transformation)
through the nodal surface of the HF wave function.

\section{Conclusions}
\label{sec:conclude}

We have devised an inhomogeneous backflow transformation for systems
consisting of electrons and either nuclei or ions represented by
pseudopotentials.  We have
applied our backflow transformation to single-determinant Slater-Jastrow
wave functions for the HEG and for atomic, molecular, and solid
systems.  In each case backflow gives a substantial reduction in the VMC
energy, and a smaller reduction in the DMC energy.

The homogeneous backflow transformation reduces the variance of the VMC
energy of the HEG by a factor of about 4, which is the largest such
factor we have encountered, and we believe that our backflow wave functions
for the HEG are very accurate.  VMC retrieves more than 99.5\% of the
DMC correlation energy in the density range studied ($r_s = 0.5$--20).
The effects of backflow on the nodes increase with $r_s$, even though
the additional percentage of the correlation energy retrieved in VMC
decreases with $r_s$, implying that the energies of dilute HEGs are
less sensitive to the nodal structure of the trial wave function than
those of denser systems.

Although backflow works very well in the HEG, as previous studies have
already concluded, we find that purely homogeneous backflow transformations
give poor results when atoms are present, as we demonstrated for
the AE lithium dimer and the AE carbon atom.  However, in these
cases inhomogeneous backflow transformations can improve the wave functions
substantially.

For the AE lithium atom the HF nodal surface of the SJ wave function
is essentially exact.  Although in this case backflow cannot improve the DMC
energy, it gives a very accurate VMC energy.  This shows that backflow
transformations can improve the wave function away from the nodes as
well as improving the nodal surface itself.  The quality of the SJ and
BF wave functions for the AE lithium dimer are much lower than for the
atom, and consequently the binding energy of the dimer is
underestimated.  The wave function and nodal surface of the AE lithium
dimer can be substantially improved by using several
determinants,\cite{bressanini_2005b} but it appears that only modest
improvements can be obtained using backflow\@.

Backflow reduces the VMC energy of the AE carbon atom by about 49\% of the
correlation energy missing at the SJ-VMC level, but at the DMC level
the improvement is smaller; the BF-DMC energy is only 18\% closer to
the exact value than SJ-DMC\@.  Backflow makes a more significant
improvement to the DMC energy of a PP
carbon atom than the AE carbon atom.  The PP and
AE carbon atoms are also cases where substantial improvements to the
wave functions can be obtained by using several determinants.  This
indicates that the SJ nodal surfaces of these two systems need a more
drastic correction than backflow transformations can provide.

When the initial nodal surface is reasonably accurate, backflow does an
excellent job in improving the VMC energy and correcting the remaining
errors in the nodal surface, as was seen in our study of the HEG and AE
lithium.  However, when the initial nodal surface is intrinsically
poor, as is the case, for example, with the HF nodal surfaces
of the carbon atom and dimer, backflow is apparently
incapable of making the gross changes to the nodal surface required to
correct the flaws, although it normally lowers the VMC and DMC
energies somewhat.  We do not believe that our backflow transformation
is capable of changing the number of nodal pockets of the starting
wave function.

The cost of using BF wave functions can be substantial, but we have
given evidence that the expense relative to that of using SJ wave functions
increases smoothly with the number of atoms in the system.  Backflow
transformations,
like Jastrow factors and unlike multideterminant expansions,
are compact parametrizations, meaning that the number of parameters
required to retrieve a given fraction of the correlation energy
increases only slowly with system size.
This can be seen by comparing the number of backflow parameters
that we have used and the energies we have obtained
for PP carbon atom, dimer, and diamond.  We
have found that it is much more efficient to move electrons one at a
time (the EBEA) than to move all the electrons at once (the CBCA), as
has been done in previous backflow calculations.  The reason for this is
that the correlation time of the energy is considerably shorter with
the EBEA\@.  It is important to use the EBEA for large systems, as the
CBCA-to-EBEA ratio of correlation times seems to increase linearly
with the number of electrons.

BF-VMC energies are normally significantly lower than SJ-VMC ones, and
therefore BF-VMC might be a useful alternative to a
(normally more expensive) SJ-DMC calculation.  The use of more
accurate trial wave functions improves the statistical efficiency of VMC and
DMC calculations.  The variance of the local energies encountered in a DMC
calculation is approximately proportional to the error in the VMC
energy, and when backflow leads to a significant reduction in the VMC energy
it also improves the statistical efficiency of DMC calculations,
even when backflow improves the DMC energy only slightly.  The improved trial
wave functions could also be useful in DMC calculations of quantities
other than the energy, which are normally more difficult to obtain
accurately than the energy.

Backflow would appear to give significant improvements in trial wave
functions for a wide variety of systems, including various different
atoms, and small and large systems.  In the present work, we have
applied the inhomogeneous backflow transformation to single-determinant
Slater-Jastrow wave functions only, but it can be combined with
multideterminant wave functions, and we will report on such
calculations elsewhere.\cite{trail_2006} It can also be combined with
pairing wave functions.\cite{lopez_rios_2006} We believe that
inhomogeneous backflow transformations will play an important role in
improving trial wave functions for use in VMC and DMC calculations.

\section{Acknowledgments}

Financial support has been provided by the Engineering and Physical
Sciences Research Council of the United Kingdom.  P.L.R.\ acknowledges
the financial support provided through the European Community's Human
Potential Programme under contract HPRN-CT-2002-00298, RTN
``Photon-Mediated Phenomena in Semiconductor Nanostructures.''
N.D.D.\ acknowledges financial support from Jesus College, Cambridge.
M.D.T.\ acknowledges financial support from the Royal Society.
Computer resources have been provided by the Cambridge-Cranfield High
Performance Computing Facility.


\appendix

\section{Constraints on the backflow parameters}


\subsection{Cusp conditions}
\label{app:cusp}

The Kato cusp conditions\cite{kato,pack} (KCC) are enforced so that
the local energy is finite when two electrons or an electron and a
nucleus are coincident.  For SJ wave functions it is common practice
to impose the electron-electron KCC (EKCC) by constraining the
parameters in the Jastrow function, and the electron-nucleus KCC
(NKCC) by constraining the orbitals in the Slater determinant. The
backflow transformation can alter the nature of the cusps, but we have
chosen to constrain the backflow parameters so that they do not modify
the KCC as applied to the Slater-Jastrow wave function.\footnote{In
principle, it would be possible to apply the KCC to the Jastrow and
backflow parameters together.  However, the resulting constraints are
configuration-dependent and involve orbital derivatives, making this
approach difficult.}

Let $i$ and $j$ be two different electrons in the system.  To satisfy
the EKCC, we require that the total backflow displacement
$\mbox{\boldmath $\xi$}_i$, has a well-defined gradient (i.e.\ it
should be cuspless)
when $r_{ij}\rightarrow 0$ if $i$ and $j$ are distinguishable
particles, and have zero gradient when $r_{ij}\rightarrow 0$ if $i$
and $j$ are indistinguishable.  Thus, the e-e term is affected by these
constraints only if $i$ and $j$ are like-spin electrons, in which case
the EKCC are satisfied if $L_{\eta}c_1=Cc_0$.

Let $I$ be a nucleus in the system.  To satisfy the NKCC, we require
that the total backflow displacement, $\mbox{\boldmath $\xi$}_i$, has a
well-defined gradient when $r_{iI}\rightarrow 0$, and that it is zero
when $r_{iI}\rightarrow 0$ if $I$ is an AE atom.  The NKCC are
satisfied if $L_{\mu,I}d_{1,I}=Cd_{0,I}$ for all $I$, and in addition,
$d_{0,I}=0$, if $I$ is an AE atom.

The constraints on the e-e-n functions, some of which only apply to
those functions centered on AE atoms, are as follows.  (We omit the
$I$ index in the parameters for clarity.)
\begin{itemize}
\item There are $3\left(N_{\rm ee}+N_{\rm en}+1\right)$ constraints
from the NKCC,
\begin{equation}
\sum_{l,m}^{(l+m=\alpha)} \left( C\varphi_{0lm} -
L_\Phi \varphi_{1lm} \right) =
\sum_{k,m}^{(k+m=\alpha)} \left( C\varphi_{k0m} -
L_\Phi \varphi_{k1m} \right) =
\sum_{k,m}^{(k+m=\alpha)} \left( C\theta_{k0m} -
L_\Phi \theta_{k1m} \right) =
0 \;\;\;\; \forall~\alpha \;.
\end{equation}
\item There are $2N_{\rm en}+1$ constraints from the EKCC,
\begin{equation}
\sum_{k,l}^{(k+l=\alpha)} \theta_{kl1} = 0 \;\;\;\; \forall~\alpha \;,
\end{equation}
and $2N_{\rm en}+1$ extra constraints for like-spin electron pairs,
\begin{equation}
\sum_{k,l}^{(k+l=\alpha)} \varphi_{kl1} = 0 \;\;\;\; \forall~\alpha \;.
\end{equation}
\item {[AE only]} There are $4\left(N_{\rm ee}+N_{\rm en}\right)+2$
constraints on $\varphi_{klm}$,
\begin{equation}
\sum_{l,m}^{(l+m=\alpha)} \varphi_{0lm} =
\sum_{l,m}^{(l+m=\alpha)} m \varphi_{0lm} =
\sum_{k,m}^{(k+m=\alpha)} \varphi_{k0m} =
\sum_{k,m}^{(k+m=\alpha)} m \varphi_{k0m} =
0 \;\;\;\; \forall~\alpha \;.
\end{equation}
\item {[AE only]} There are $3\left(N_{\rm ee}+N_{\rm en}\right)+2$
constraints on $\theta_{klm}$,
\begin{equation}
\sum_{l,m}^{(l+m=\alpha)} \theta_{0lm} =
\sum_{l,m}^{(l+m=\alpha)} m \theta_{0lm} =
\sum_{k,m}^{(k+m=\alpha)} m \theta_{k0m} =
0 \;\;\;\; \forall~\alpha \;.
\end{equation}
\end{itemize}

These constraints form an indeterminate system of homogeneous
algebraic linear equations for the e-e-n parameters.
Hence, a subset of the parameters can be put in terms of the rest.
This subset can be determined from the ``free'' parameters by
putting the constraints in matrix form and using Gaussian
elimination.  This procedure is the one described in
Ref.\ \onlinecite{drummond_2004}, where it is applied to the
parameters in the e-e-n term of the Jastrow factor.


\subsection{Constraints for irrotational backflow}
\label{app:nocurl}

In the derivation of homogeneous backflow in
Ref.\ \onlinecite{holzmann_2003} it was suggested that the backflow
displacement should satisfy $\mbox{\boldmath $\xi$}_i=\nabla_{i}Y$, where
$Y=Y({\bf R})$ is an object called the \textit{backflow potential}.
This equation is already satisfied by both the e-e and e-n terms, by
definition, and it can be imposed on the e-e-n functions by using an
appropriate set of constraints.  From
$\nabla_i \times \mbox{\boldmath $\xi$}_i = {\bf 0}$, it follows that
\begin{equation}
r_{ij} \frac{\partial}{\partial r_{iI}}
\left[ \Phi_i^{jI} f(r_{iI};L_{\Phi,I}) \right] =
r_{iI} \frac{\partial}{\partial r_{ij}}
\left[ \Theta_i^{jI} f(r_{iI};L_{\Phi,I}) \right] \;,
\end{equation}
for all $i$, $j$, and $I$, and all $r_{ij}$, $r_{iI}$, and $r_{jI}$.
For $C>0$, this results in the equation
\begin{equation}
\label{eq:no_curl_C_gt_0}
\left( C+k \right) \varphi_{k,l,m-1} -
L_\Phi \left( k+1 \right) \varphi_{k+1,l,m-1} -
\left( m+1 \right) \theta_{k-2,l,m+1}+
L_\Phi \left( m+1 \right) \theta_{k-1,l,m+1} =0 \;,
\end{equation}
while for $C=0$,
\begin{equation}
\left( k+1 \right) \varphi_{k+1,l,m-1} -
\left( m+1 \right) \theta_{k-1,l,m+1} =0 \;.
\end{equation}
In both cases, $0\leq k\leq N_{\rm en}+2$, $0\leq l\leq N_{\rm en}$,
and $0\leq m\leq N_{\rm ee}+1$, and parameters with indices out of the
allowed range are to be taken as equal to zero.  The $I$ index has
been omitted for clarity.

The application of these constraints results in a reduction in the
number of free parameters by more than one half, as one would expect,
because an equivalent backflow displacement would be obtained by
parameterizing the scalar field $Y$ and computing its gradient, whereas
we use \textit{two} scalar fields in the full e-e-n term.


\section{Zeroing the backflow displacement at AE atoms}
\label{app:zero_ae}

When AE atoms are present, the NKCC cannot be fulfilled unless the
backflow displacement at the nuclear position is zero.  This can be
obtained by applying smooth cutoffs around such atoms.
In this scheme, an artificial multiplicative cutoff function
$g(r_{iI})$ is applied to all contributions to the backflow
displacement of particle $i$ that do not depend on the distance
$r_{iI}$ to the AE atom $I$.  This includes the homogeneous backflow
displacement and the inhomogeneous contributions centered on each atom
$J\neq I$.

The $g(r_{iI})$ function must go to zero at $r_{iI}\rightarrow 0$ and
become unity when $r_{iI}$ is equal to or greater than a threshold
$L_{g,I}$.  For the local energy to be well-defined, we require that
$g(r_{iI})$ and its first two derivatives be continuous at
$r_{iI}=L_{g,I}$, and to fulfill the NKCC correctly, $g(r_{iI})$ and
its first derivative must go to zero at $r_{iI}=0$.  The simplest
$g(r_{iI})$ obeying these conditions is the fourth-order polynomial,
\begin{equation}
g(r_{iI}) = \left( \frac{r_{iI}} {L_{g,I}} \right)^2 \left[ 6 - 8
\left( \frac{r_{iI}} {L_{g,I}} \right) + 3 \left( \frac{r_{iI}}
{L_{g,I}} \right)^2 \right] \;,
\end{equation}
which we have used in our calculations.  Although it is perfectly
possible to optimize the $L_{g,I}$, we have used fixed values for
simplicity: $1$~a.u.\ in the AE atoms and about half the interatomic
distance in the AE ${\rm Li}_2$ molecule.



\begin{thebibliography}{99}


\bibitem{ceperley_1980} D.~M.\ Ceperley and B.~J.\ Alder, Phys.\ Rev.\
Lett.\ \textbf{45}, 566 (1980).

\bibitem{foulkes_2001} W.~M.~C.\ Foulkes, L.\ Mit\'as, R.~J.\ Needs,
and G.\ Rajagopal, Rev.\ Mod.\ Phys.\ \textbf{73}, 33 (2001).

\bibitem{anderson_1975} J.~B.\ Anderson, J.\ Chem.\ Phys.\
\textbf{63}, 1499 (1975).

\bibitem{shull_1959} Two-electron pairing functions were named
  ``geminals'' by H.\ Shull, J.\ Chem.\ Phys.\ \textbf{30}, 1405
  (1959), to distinguish them from one-electron orbitals.

\bibitem{fock_1950} V.~A.\ Fock, Dokl.\ Akad.\ Nauk.\ SSSR\
  \textbf{73}, 735 (1950).

\bibitem{hurley_1953} A.~C.\ Hurley, J.\ Lennard-Jones, and J.~A.\
Pople, Proc.\ Roy.\ Soc.\ (London) \textbf{A220}, 446 (1953).

\bibitem{casula_2003} M.\ Casula and S.\ Sorella, J.\ Chem.\ Phys.\
\textbf{119}, 6500 (2003).

\bibitem{casula_2004} M.\ Casula, C.\ Attaccalite, and S.\ Sorella,
J.\ Chem.\ Phys.\ \textbf{121}, 7110 (2004).

\bibitem{bouchaud_1987} J.~P.\ Bouchaud and C.\ Lhuillier, Europhys.\
Lett.\ \textbf{3}, 1273 (1987).

\bibitem{bajdich_2006} M.\ Bajdich, L.\ Mitas, G.\ Drobny, L.~K.\
Wagner, and K.~E.\ Schmidt, Phys.\ Rev.\ Lett.\ \textbf{96}, 130201
(2006).

\bibitem{wigner_1934} E.\ Wigner and F.\ Seitz, Phys.\ Rev.\
\textbf{46}, 509 (1934).

\bibitem{feynman_1954} R.~P.\ Feynman, Phys.\ Rev.\ \textbf{94}, 262
  (1954).

\bibitem{feynman_1956} R.~P.\ Feynman and M.\ Cohen, Phys.\ Rev.\
\textbf{102}, 1189 (1956).

\bibitem{pandharipande_1973} V.~R.\ Pandharipande and N.\ Itoh, Phys.\
Rev.\ A\ \textbf{8}, 2564 (1973).

\bibitem{schmidt_1979} K.~E.\ Schmidt and V.~R.\ Pandharipande, Phys.\
Rev.\ B\ \textbf{19}, 2504 (1979).

\bibitem{manousakis_1983} E.\ Manousakis, S.\ Fantoni, V.~R.\
Pandharipande, and Q.~N.\ Usmani, Phys.\ Rev.\ B\ \textbf{28}, 3770
(1983).

\bibitem{lee_1981} M.~A.\ Lee, K.~E.\ Schmidt, M.~H.\ Kalos, and
G.~V.\ Chester, Phys.\ Rev.\ Lett.\ \textbf{46}, 728 (1981).

\bibitem{kwon_1993} Y.\ Kwon, D.~M.\ Ceperley, and R.~M.\ Martin,
Phys.\ Rev.\ B\ \textbf{48}, 12037 (1993).

\bibitem{kwon_1998} Y.\ Kwon, D.~M.\ Ceperley, and R.~M.\ Martin,
Phys.\ Rev.\ B\ \textbf{58}, 6800 (1998).

\bibitem{zong_2002} F.~H.\ Zong, C.\ Lin, and D.~M.\ Ceperley, Phys.\
Rev.\ E\ \textbf{66}, 036703 (2002).

\bibitem{holzmann_2003} M.\ Holzmann, D.~M.\ Ceperley, C.\ Pierleoni,
and K.\ Esler, Phys.\ Rev.\ E\ \textbf{68}, 046707 (2003).

\bibitem{pierleoni_2004} C.\ Pierleoni, D.~M.\ Ceperley, and M.\
Holzmann, Phys.\ Rev.\ Lett.\ \textbf{93}, 146402 (2004).

\bibitem{schmidt_1981} K.~E.\ Schmidt, M.~A.\ Lee, M.~H.\ Kalos, and
G.~V.\ Chester, Phys.\ Rev.\ Lett.\ \textbf{47}, 807 (1981).

\bibitem{drummond_2004} N.~D.\ Drummond, M.~D.\ Towler, and R.~J.\
Needs, Phys.\ Rev.\ B\ \textbf{70}, 235119 (2004).

\bibitem{casino} R.~J.\ Needs, M.~D.\ Towler, N.~D.\ Drummond, and P.\
L\'opez R\'{\i}os, \textsc{casino} User's guide Version 2.0.0 (2006).

\bibitem{umrigar_1993} C.~J.\ Umrigar, M.~P.\ Nightingale, and K.~J.\
Runge, J.\ Chem.\ Phys.\ \textbf{99}, 2865 (1993).

\bibitem{umrigar_1988} C.~J.\ Umrigar, K.~G.\ Wilson, and J.~W.\
Wilkins, Phys.\ Rev.\ Lett.\ \textbf{60}, 1719 (1988).

\bibitem{kent_1999} P.~R.~C.\ Kent, R.~J.\ Needs, and G.\ Rajagopal,
Phys.\ Rev.\ B\ \textbf{59}, 12344 (1999).

\bibitem{drummond_2005} N.~D.\ Drummond and R.~J.\ Needs, Phys.\ Rev.\
B\ \textbf{72}, 085124 (2005).

\bibitem{crystal98} V.~R.\ Saunders, R.\ Dovesi, C.\ Roetti, M.\
Caus\`a, N.~M.\ Harrison, R.\ Orlando, and C.~M.\ Zicovich-Wilson,
\textsc{crystal98} User's Manual, Universit\`a di Torino, Torino
(1998).

\bibitem{ma_2005} A.\ Ma, N.~D.\ Drummond, M.~D.\ Towler, and R.~J.\
Needs, J.\ Chem.\ Phys.\ \textbf{122}, 224322 (2005).

\bibitem{trail_ppots1} J.~R.\ Trail and R.~J.\ Needs, J.\ Chem.\
Phys.\ \textbf{122}, 014112 (2005).

\bibitem{trail_ppots2} J.~R.\ Trail and R.~J.\ Needs, J.\ Chem.\
Phys.\ \textbf{122}, 174109 (2005).

\bibitem{mitas_1991} L.\ Mitas, E.~L.\ Shirley, and D.~M.\ Ceperley,
J.\ Chem.\ Phys.\ \textbf{95}, 3467 (1991).

\bibitem{segall_2002} M.~D.\ Segall, P.~J.~D.\ Lindan, M.~I.~J.\ Probert,
C.~J.\ Pickard, P.~J.\ Hasnip, S.~J.\ Clark, and M.~C.\ Payne, J.\
Phys.\ Condens.\ Matt.\ \textbf{14}, 2717 (2002).

\bibitem{perdew_1996} J.~P.\ Perdew, K.\ Burke, and M.\ Ernzerhof,
Phys.\ Rev.\ Lett.\ \textbf{77}, 3865 (1996).

\bibitem{alfe_2004} D.\ Alf\`e and M.~J.\ Gillan, Phys.\ Rev.\ B\
\textbf{70}, 161101(R) (2004).

\bibitem{dolg_1996} M.\ Dolg, Chem.\ Phys.\ Lett.\ \textbf{250}, 75 (1996).

\bibitem{bressanini_2002} D.\ Bressanini, D.~M.\ Ceperley, and P.~J.\
Reynolds, in \textit{Recent Advances in Quantum Monte Carlo Methods},
edited by W.~A.\ Lester, Jr., S.~M.\ Rothstein, and S.\ Tanaka (World
Scientific, Singapore, 2002), 2nd ed.

\bibitem{davidson_1991} E.~R.\ Davidson, S.~A.\ Hagstrom, S.~J.\
Chakravorty, V.~M.\ Umar, and C.~F.\ Fischer, Phys.\ Rev.\ A\
\textbf{44}, 7071 (1991).

\bibitem{chakravorty_1993} S.~J.\ Chakravorty, S.~R.\ Gwaltney, E.~R.\
Davidson, F.~A.\ Parpia, and C.~F.\ Fischer, Phys.\ Rev.\ A\
\textbf{47}, 3649 (1993).

\bibitem{cade_1974} P.~E.\ Cade and A.~C.\ Wahl, At.\ Data\ Nucl.\
Data\ Tables\ \textbf{13}, 340 (1974).

\bibitem{bressanini_2005b} D.\ Bressanini, G.\ Morosi, and S.\
Tarasco, J.\ Chem.\ Phys.\ \textbf{123}, 204109 (2005).

\bibitem{filippi_1996} C.\ Filippi and C.~J.\ Umrigar, J.\ Chem.\
Phys.\ \textbf{105}, 213 (1996).

\bibitem{barnett_2000} R.~N.\ Barnett, Z.\ Sun, and W.~A.\ Lester,
Jr., J.\ Chem.\ Phys.\ \textbf{114}, 2013 (2000).

\bibitem{glauser_1992} W.~A.\ Glauser, W.~R.\ Brown, W.~A.\ Lester,
Jr., D.\ Bressanini, B.~L.\ Hammond, and M.~L.\ Koszykowski, J.\
Chem.\ Phys.\ \textbf{97}, 9200 (1992).

\bibitem{sato_2002} T.\ Sato, K.\ Ohashi, T.\ Sudoh, K.\ Haruna, and
H.\ Maeta, Phys.\ Rev.\ B \textbf{65}, 092102 (2002).

\bibitem{flyvbjerg_1989} H.~Flyvbjerg and H.~G.~Petersen, J.\ Chem.\
Phys.\ \textbf{91}, 461 (1989).

\bibitem{ceperley_1986} D.~M.\ Ceperley, J.\ Stat.\ Phys.\
\textbf{43}, 815 (1986).

\bibitem{ma_2005b} A.\ Ma, N.~D.\ Drummond, M.~D.\ Towler, and R.~J.\
Needs, Phys.\ Rev.\ E\ \textbf{71}, 066704 (2005).

\bibitem{bressanini_2005a} D.\ Bressanini and P.~J.\ Reynolds, Phys.\
Rev.\ Lett.\ \textbf{95}, 110201 (2005).

\bibitem{bajdich_2005} M.\ Bajdich, L.\ Mitas, G.\ Drobny, and L.~K.\
Wagner, Phys.\ Rev.\ B\ \textbf{72}, 075131 (2005).

\bibitem{trail_2006} J.~R.\ Trail \textit{et al.}, unpublished.

\bibitem{lopez_rios_2006} P.\ L\'opez R\'{\i}os \textit{et al.},
unpublished.

\bibitem{kato} T.\ Kato, Commun.\ Pure\ Appl.\ Math.\ \textbf{10}, 151
  (1957).

\bibitem{pack} R.~T.\ Pack and W.~B.\ Brown, J.\ Chem.\ Phys.\
  \textbf{45}, 556 (1966).

\end{thebibliography}
\end{document}